\documentclass[pdflatex,sn-mathphys-ay]{sn-jnl}

\usepackage[T1]{fontenc}
\usepackage{amsmath,amssymb,amsfonts}
\usepackage{booktabs}
\usepackage{multirow}
\usepackage{enumitem}
\usepackage{tabularx}
\usepackage{array}
\usepackage{float}
\usepackage{placeins}
\usepackage{longtable}
\usepackage{tikz}
\usetikzlibrary{positioning,shapes.geometric,arrows.meta,fit,backgrounds,calc}
\usepackage{orcidlink}

\definecolor{tabBlue}  {HTML}{1F77B4}
\definecolor{tabOrange}{HTML}{FF7F0E}
\definecolor{tabGreen} {HTML}{2CA02C}
\definecolor{tabRed}   {HTML}{D62728}
\definecolor{tabGray}  {HTML}{7F7F7F}

\newcommand{\ecmcontract}{\mathcal{C}}

\usepackage{microtype}
\begin{document}

\title[ECM Contracts for Embodied Agents]{\texorpdfstring{ECM Contracts: Contract-Aware, Versioned,\\ and Governable Capability Interfaces\\ for Embodied Agents}{ECM Contracts: Contract-Aware, Versioned, and Governable Capability Interfaces for Embodied Agents}}


\author[1]{\fnm{Xue} \sur{Qin}\,\orcidlink{0009-0009-3642-2663}}\email{qinxue@me.com}

\author[2]{\fnm{Simin} \sur{Luan}\,\orcidlink{0000-0003-1138-1892}}\email{luansiminiot@gmail.com}

\author*[3]{\fnm{Cong} \sur{Yang}\,\orcidlink{0000-0002-8314-0935}}\email{cong.yang@suda.edu.cn}

\author*[2]{\fnm{Zhijun} \sur{Li}\,\orcidlink{0000-0001-9129-9957}}\email{lizhijun\_os@hit.edu.cn}

\affil[1]{\orgdiv{School of Software}, \orgname{Harbin Institute of Technology}, \orgaddress{\city{Harbin}, \country{China}}}

\affil[2]{\orgdiv{School of Computer Science and Technology}, \orgname{Harbin Institute of Technology}, \orgaddress{\city{Harbin}, \country{China}}}

\affil[3]{\orgdiv{School of Future Science and Engineering}, \orgname{Soochow University}, \orgaddress{\city{Suzhou}, \country{China}}}

\abstract{Embodied robots are increasingly assembled at runtime from independently published capability modules, yet their interfaces are specified only as message types, so failures such as frame mismatches, resource conflicts, and version skew surface only at execution. We present \emph{ECM Contracts}, a contract-based interface model that augments a module's functional signature with five further dimensions of embodied execution (behavioral assumptions, resource requirements, permission boundaries, recovery semantics, and version compatibility), together with a compatibility framework that decides composition, upgrade, and release before deployment. We evaluate the prototype by predicting integration failures that third parties documented independently, on two substrates: the ROBUST corpus of reproduced ROS bugs and a provenance-screened corpus of 16 documented ROS\,2 bugs. Contracts are reconstructed from each module's published interface and scored by a checker frozen by content hash before reconstruction. Contract checking predicts 31\% and 62\% of these failures against at most 19\% for the strongest type and quality-of-service baselines, significant on both substrates (exact McNemar $p=0.0078$ and $p=0.016$), with no false positives on negative controls and three predictions confirmed live in ROS\,2. An independent, bug-blind re-derivation of every scored win from the public artefacts corroborated the retained wins and corrected three of our own cases downward. We report each dimension at the tier its evidence supports: resource requirements and named-entity version changes form a validated core that every baseline misses, the behavioral dimension is largely subsumed by middleware quality-of-service, and permission and recovery are forward-looking.}


\maketitle

\section{Introduction}
\label{sec:introduction}

Consider a service robot assembled from three independently published capability modules: navigation, grasping, and human handover. At the level of message types the modules appear composable, so a schema-based check accepts the assembly. The deployed system fails anyway. The grasp module holds the object in a base-link frame while the handover module expects it in a camera frame, so the transfer pose is computed in the wrong reference frame. The upgraded navigation module assumes a higher localization refresh rate than the runtime delivers, so the planner acts on stale poses. The handover module claims close-range human-interaction permissions that were never granted at installation, so a safety gate blocks it at the worst moment. None of these faults is visible from input and output signatures alone, and each surfaces only when the modules execute together on a physical robot.

\begin{figure}[!ht]
\centering
\includegraphics[width=0.85\textwidth]{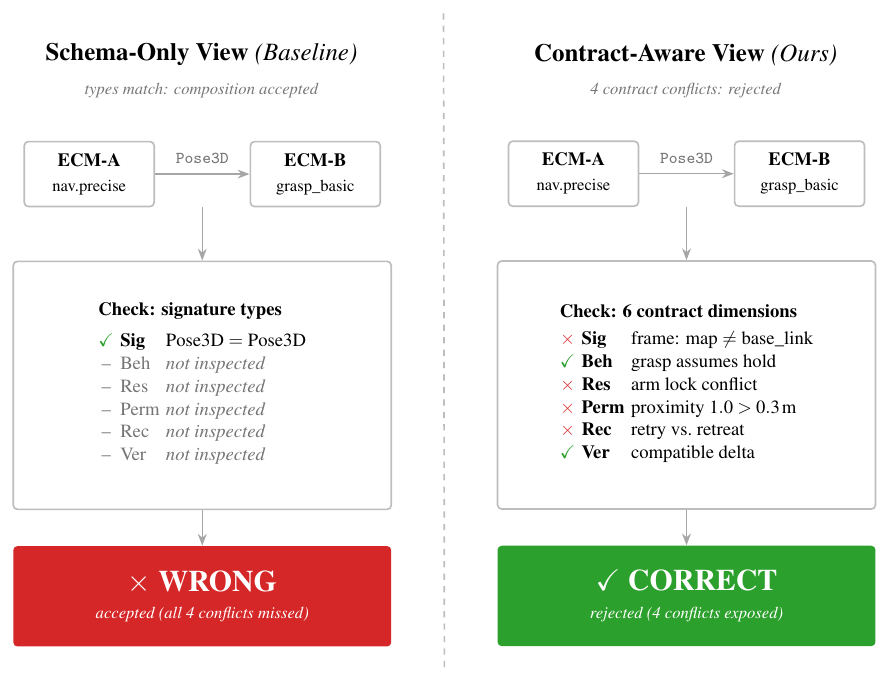}
\caption{Schema-only checking (left) sees only type compatibility between two ECMs and accepts the composition. Contract-aware checking (right) evaluates the full contract and reveals hidden incompatibilities (frame mismatch, resource conflict, permission gap, and version skew) that would cause runtime failure but are invisible to signature matching alone.}
\label{fig:hero}
\end{figure}

These are not contrived examples. Frame-of-reference mistakes of exactly this kind are common enough that an automated checker, \textsc{PhysFrame}, found 190 frame inconsistencies across 180 public ROS projects~\citep{kate2021physframe}. Capability and resource mismatches, where a module assumes hardware or software a platform does not provide, motivated semantic robot description languages that reason about capability dependencies before execution~\citep{kunze2011srdl}, and the difficulty of composing kinematic and interface models across modules without hidden assumptions has driven a line of work on composable domain-specific models~\citep{schneider2023composable}. Two large empirical studies of real ROS systems make the prevalence concrete: the \textsc{Robust} dataset catalogues 221 reproduced bugs across seven robotics systems~\citep{timperley2024robust}, and a study of ROS misconfigurations characterises faults that arise precisely when components make conflicting assumptions about the robot they share~\citep{canelas2024misconfig}. Failure classes that recur across these studies include frame mismatch, resource and capability conflict, and version skew between an upgraded module and the rest of the system.

The existing responses are point solutions. \textsc{PhysFrame} checks one property, physical frames~\citep{kate2021physframe}; \textsc{Srdl} models capability dependencies~\citep{kunze2011srdl}; composable-model DSLs target interface and kinematics modelling~\citep{schneider2023composable}; and the empirical studies document the failures without offering a mechanism to prevent them~\citep{timperley2024robust,canelas2024misconfig}. Each addresses a single facet of the problem, and none gives a developer a unified, machine-checkable declaration that can be evaluated before deployment to decide whether two modules compose, whether an upgrade is safe, and whether a release is ready. The field needs an interface description for embodied modules that spans the dimensions on which real integrations break, evaluated as one pre-deployment check rather than one property at a time.

We present \emph{ECM Contracts}, a contract-based interface model that extends a module's signature with the behavioral, resource, permission, recovery, and version information needed to decide composition, upgrade, and release compatibility before a robot runs. A single contract model grounds a compatibility framework, which in turn grounds a release discipline, and all three are realized by a reference prototype that we evaluate against real documented bugs and in a real runtime. Our contributions are as follows:
\begin{itemize}[leftmargin=*]
    \item \textbf{ECM contract model.} A contract structure for embodied capability modules that augments the functional signature with behavioral assumptions, resource requirements, permission boundaries, recovery semantics, and version compatibility, together with the compatibility operators that decide installation, composition, and upgrade (Sections~\ref{sec:contract}--\ref{sec:compatibility}).
    \item \textbf{Release discipline.} An embodied semantic-versioning scheme with compatibility classes, deprecation and migration rules, and policy-gated upgrade checks, so that module evolution is governed rather than ad hoc (Section~\ref{sec:release}).
    \item \textbf{Evidence on real documented bugs.} We reconstruct module contracts from published interfaces and a frozen checker, then test whether they predict third-party-documented integration bugs on two independent substrates. ECM Contracts recall 31\% (10 of 32) of integration bugs on the \textsc{Robust} corpus and 62\% (10 of 16) on a ROS2 bug corpus, against at most 19\% for schema-only, schema-plus-QoS, and semver-only checking, while accepting every matched-good composition (0\% false positives on negative controls). Three predicted failures, a quality-of-service mismatch, a resource-ownership conflict in the \texttt{ros2\_control} resource manager, and a coordinate-frame mismatch through the \texttt{tf2} stack, reproduce as execution events in a real ROS\,2 runtime (Section~\ref{sec:evaluation}).
    \item \textbf{A blind reconstruction study of interface-derivability.} To separate the contract model's predictive value from the risk that the authors encoded the documented failure, we ran a blind reconstruction study: independent annotators, given only each module's declared interface with the bug narrative removed, judged whether a composition would fail and on which dimension. The signature, resource, and behavioral conflicts were independently reproduced from the interface facts, and matched-good controls were cleared, which indicates these conflicts are interface-derivable (Section~\ref{sec:eval_blind}).
    \item \textbf{An account of which dimensions carry the evidence.} The resource dimension is the cleanest win, catching contention conflicts that every baseline misses. The version dimension is restricted to documented named-entity interface changes after a hash-versus-semver heuristic proved unreliable. The signature dimension is partly co-detected by a schema baseline; the behavioral dimension is largely subsumed by QoS-aware schema checking; and the permission and recovery dimensions are forward-looking, grounded in safety-standard mechanisms but not yet backed by documented composition bugs (Sections~\ref{sec:evaluation}--\ref{sec:discussion}).
\end{itemize}

This work completes a layer in a broader research program on modular embodied intelligence. Prior papers introduced the ECM concept~\citep{qin2025ecm}, continuous capability evolution~\citep{qin2025evolution}, runtime governance~\citep{qin2025governance}, and upgrade governance~\citep{qin2026upgrade}. The present paper adds the \emph{contract substrate}: the formal basis for deciding when composition is safe, when an upgrade is compatible, and when a release is ready, validated against bugs that real robotics systems have actually reported.
\subsection{Motivation and Problem Formulation}
\label{sec:motivation}

Before formalizing the ECM Contract model, we ground our approach in two motivating scenarios that illustrate why conventional interface descriptions are insufficient for embodied capability composition and evolution. We then distill three formal research questions that structure the remainder of the paper.

\subsubsection{Motivating Scenario 1: Composition Failure}

Consider a service robot tasked with delivering a medical instrument from a storage shelf to a clinician. The task planner decomposes this into three ECMs:
\begin{itemize}[leftmargin=*]
    \item \textbf{ECM-A (navigate-to-shelf):} Navigates the mobile base to the storage location using a 2D laser-based SLAM module, expressing the goal pose in a map-frame coordinate system.
    \item \textbf{ECM-B (grasp-instrument):} Activates a 6-DOF arm to grasp the target object, expecting a wrist-mounted RGB-D camera and force-torque sensor, with the grasp pose expressed in the end-effector frame.
    \item \textbf{ECM-C (handover-to-clinician):} Extends the arm and releases the object into the clinician's hand, requiring human-proximity permission and a compliant control mode for safe physical interaction.
\end{itemize}

At the schema level, these three modules chain correctly: ECM-A produces a robot pose, ECM-B accepts a robot pose and produces a grasp state, ECM-C accepts a grasp state and produces a handover confirmation. A schema-only composition checker would approve this pipeline. However, at the embodied execution level, multiple hidden incompatibilities exist:
\begin{itemize}[leftmargin=*]
    \item \textbf{Frame mismatch.} ECM-A operates in a 2D map frame, while ECM-B expects a 3D base-link frame. Without explicit frame transformation, the grasp target localization will fail.
    \item \textbf{Resource conflict.} ECM-B requires a wrist-mounted force-torque sensor for grasp force feedback, but the specific robot platform has only a fingertip pressure sensor, which provides different data semantics and resolution.
    \item \textbf{Permission gap.} ECM-C requires human-proximity permission (the robot must be authorized to operate within 0.5 meters of a person), but the runtime's current policy grants only corridor-navigation permission, not close-interaction permission.
    \item \textbf{Recovery conflict.} ECM-B's recovery strategy upon grasp failure is to retry with increased force, while ECM-A's post-failure assumption is that the robot should retreat to a safe waypoint. If ECM-B fails and retries, the navigation module's state machine enters an inconsistent state.
\end{itemize}

Each of these incompatibilities arises from dimensions that conventional software interfaces do not capture: coordinate frame semantics, hardware resource specifications, policy-level permissions, and failure recovery protocols.

\subsubsection{Motivating Scenario 2: Upgrade Failure}

A second class of failures emerges during capability evolution. Suppose that the grasp module (ECM-B) is upgraded from version 1.0 to version 2.0. The new version introduces a learning-based grasp planner with the following changes:
\begin{itemize}[leftmargin=*]
    \item The input schema remains identical: the module still accepts a target object pose and a robot configuration.
    \item The module now requires GPU access for real-time inference, whereas version 1.0 used only CPU-based heuristics.
    \item The failure-recovery semantics have changed: version 2.0 attempts a different grasp pose on failure (explore-and-retry), whereas version 1.0 retried the same pose with increased force.
    \item The module now emits a confidence score alongside the grasp result, which downstream modules may or may not consume.
\end{itemize}

From a conventional semantic versioning perspective, the schema has not changed, so this might be classified as a minor or even patch release. But from an embodied execution perspective, the upgrade is breaking: the planner's resource allocation does not provision GPU, the recovery coordinator's assumption about retry behavior is invalidated, and the policy engine has not reviewed the new model's safety properties. The result is a deployment that passes all schema checks but fails in production, precisely the category of failure that a contract-aware system should prevent.

\subsubsection{Problem Formulation}

These scenarios motivate three formal research questions:

\noindent\textbf{RQ1:} \emph{What constitutes an ECM interface for embodied execution?} Traditional software interfaces describe functional signatures. We ask what additional dimensions must be captured to fully specify the assumptions and guarantees of an embodied capability module.

\noindent\textbf{RQ2:} \emph{When are two or more ECMs safely composable?} Given a proposed composition of ECMs, we ask what conditions must hold (across types, behaviors, resources, permissions, and recovery semantics) for the composition to be considered safe.

\noindent\textbf{RQ3:} \emph{When is an upgraded ECM release-compatible with existing runtimes and policies?} Given a new version of an ECM, we ask what contract-level checks must be satisfied for the upgrade to be deployed without breaking existing compositions, planner assumptions, or governance policies.

\section{Related Work}
\label{sec:related}

ECM Contracts sit between three lines of prior work: runtime verification and behavioral contracts for ROS, interface and composition modeling, and point solutions for individual interface concerns. We position the contract model against each and summarize the comparison as a feature matrix in Table~\ref{tab:feature_matrix}.

\noindent\textbf{Behavioral Contracts and Runtime Verification.} The closest work attaches assume-guarantee contracts to ROS nodes and compiles them into runtime monitors. The ROS Contract Language and its Vanda tool express per-node first-order assumptions and guarantees and derive system-level properties through a composition calculus~\citep{luckcuck2023compositional}. ROSMonitoring instruments the message stream and checks temporal properties over execution traces~\citep{ferrando2020rosmonitoring}. These approaches reason about behavior, and they observe it at runtime, after deployment. ECM Contracts instead check installation, composition, and upgrade before execution, and they cover dimensions a behavioral monitor does not express, including exclusive resource ownership and cross-version interface compatibility.

\noindent\textbf{Interface and Composition Modeling.} Composable domain-specific models capture kinematic and interface structure so that modules can be combined without hidden assumptions~\citep{schneider2023composable}. These models are strong on the signature dimension and on design-time structure, and they do not address behavioral QoS, resource contention, or version evolution. ECM Contracts reuse the same idea of an explicit interface description and extend it with the runtime-facing dimensions that the documented failures in Section~\ref{sec:evaluation} require.

\noindent\textbf{Point Solutions for Single Concerns.} Several tools each target one interface concern. PhysFrame type-checks coordinate frames and found 190 inconsistencies across 180 ROS projects~\citep{kate2021physframe}. Semantic robot description languages reason about whether a platform has the capabilities and resources an action needs~\citep{kunze2011srdl}. SROS2 enforces publish and subscribe permissions through access-control policies~\citep{mayoralvilches2022sros2}. Each tool is effective within its concern. None checks the others, and none gives a single pre-deployment verdict over signature, behavior, resource, permission, recovery, and version together.

\noindent\textbf{Empirical Grounding.} Several studies of real systems establish that these failures are frequent rather than hypothetical. The ROBUST dataset reproduces 221 bugs across seven robotics systems~\citep{timperley2024robust}, and a study of ROS misconfigurations characterizes the cross-component assumption mismatches that arise at integration time~\citep{canelas2024misconfig}. Beyond robotics, configuration errors, including interface-level parameter mismatches between components, are a major source of system failures~\citep{yin2011config}; about one in seven ROS bugs stems from dependency misuse~\citep{fischernielsen2020dependency}; and a cross-system characterization of cyber-physical software finds interface faults prevalent across drones, automotive, and robotics projects~\citep{zampetti2022cps}. Our evaluation reuses these independent sources as ground truth rather than constructing its own.

\noindent\textbf{Interface Evolution and Breaking Changes.} The version dimension connects to a body of empirical work on how interfaces break across releases. Studies of large library ecosystems find that breaking changes are common and that version numbers signal them poorly: many Java library upgrades break binary compatibility~\citep{jezek2015break}, semantic-versioning rules are frequently violated in Maven Central while most clients remain unaffected~\citep{ochoa2022semver}, and breaking changes reach downstream npm clients even through minor and patch releases~\citep{venturini2023broke}. Developers also break interfaces deliberately, for reasons the version number does not record~\citep{brito2018break}. ECM Contracts address the same failure class in the robotics setting with one difference in method: these studies measure breakage after the fact by mining build or test failures, whereas the version dimension records the named-entity change in the interface so that an affected consumer is flagged before deployment. That version numbers under-signal breakage is also why our semver-only baseline is weak by construction rather than by an unfair comparison.

\noindent\textbf{Validating Predictors on Documented Failures.} Our evaluation method follows an empirical-software-engineering tradition of measuring a detector by its recall on real, independently documented defects rather than on a benchmark its own authors built. In that setting static bug detectors find only a small fraction of real bugs~\citep{habib2018howmany}, their real-world effectiveness is modest even for a narrow defect class~\citep{tomassi2021realworld}, and the quality of the bug dataset itself changes the measured numbers~\citep{croft2023dataquality}. These findings motivate the discipline we adopt: a checker frozen by content hash before scoring, third-party bugs as ground truth, a gold-dimension crosswalk fixed independently of the checker, and recall reported with confidence intervals. The contract literature supplies the complementary motivation, since empirical studies of deployed code find that contracts, where present, are simple and under-used~\citep{estler2014contracts,dietrich2017wild}, and that stronger specifications catch faults developers miss once someone writes them~\citep{polikarpova2013specs}. An explicit, machine-checked contract layer is meant to close that gap for embodied modules.

\begin{table}[!ht]
\centering
\caption{Coverage of Existing Approaches Across the Six Contract Dimensions. Y = covered, P = partial, --- = not addressed. ``Pre-deploy'' marks methods that decide before execution; ``Code'' marks publicly available implementations.}
\label{tab:feature_matrix}
\small
\begin{tabularx}{\linewidth}{@{}>{\raggedright\arraybackslash}X*{6}{c}cc@{}}
\toprule
\textbf{Approach} & \textbf{Sig} & \textbf{Beh} & \textbf{Res} & \textbf{Perm} & \textbf{Rec} & \textbf{Ver} & \textbf{Pre-deploy} & \textbf{Code} \\
\midrule
ECM Contracts (this work) & Y & P & Y & P & P & Y & Y & Y \\
RCL / Vanda~\citep{luckcuck2023compositional} & P & Y & --- & P & --- & --- & --- & --- \\
ROSMonitoring~\citep{ferrando2020rosmonitoring} & P & Y & --- & --- & --- & --- & --- & Y \\
ROS~2 middleware (type + QoS) & Y & P & --- & --- & --- & --- & Y & Y \\
SROS2~\citep{mayoralvilches2022sros2} & --- & --- & P & Y & --- & --- & Y & Y \\
Composable models~\citep{schneider2023composable} & Y & --- & --- & --- & --- & P & Y & P \\
PhysFrame~\citep{kate2021physframe} & Y & --- & --- & --- & --- & --- & Y & Y \\
\bottomrule
\end{tabularx}
\end{table}

Table~\ref{tab:feature_matrix} shows the gap that motivates a multi-dimensional contract. Existing tools cover the signature and behavior dimensions well, and behavioral QoS compatibility is already enforced by the ROS~2 middleware at bring-up. No existing tool addresses exclusive resource ownership, recovery semantics, and cross-version compatibility together, and none gives a single pre-deployment verdict over all six dimensions. Our evaluation (Section~\ref{sec:evaluation}) measures where this breadth pays off: the resource and version dimensions predict real failures that the strongest available baselines miss entirely, while the behavioral dimension adds little beyond the QoS check the middleware already provides.

\section{The ECM Contract Framework}
\label{sec:framework}

This section presents the ECM contract framework, illustrated in Figure~\ref{fig:hero}. We first bound the class of systems the framework targets (Section~\ref{sec:scope}), then define the six-dimension contract that specifies a module's embodied interface (Section~\ref{sec:contract}) and the compatibility operators that decide installation, composition, and upgrade (Section~\ref{sec:compatibility}). We then give the release discipline that governs how modules evolve (Section~\ref{sec:release}) and the prototype that implements the checker (Section~\ref{sec:prototype}).

\subsection{Scope and System Model}
\label{sec:scope}

ECM Contracts target the modular-node ecosystem built on the Robot Operating System, in both its ROS~1 and ROS~2 forms~\citep{quigley2009,macenski2022}. In this ecosystem a working robot is assembled from independently developed packages that communicate through typed topics and services, share coordinate frames through a transform tree, and are versioned and distributed as separate units. The embodiment class we address is the mobile bases, manipulators, and service robots composed this way: the differential-drive bases, articulated arms, and human-facing service platforms exemplified in our evaluation by the kobuki base, the Universal Robots and Motoman arm drivers, the mavros/PX4 flight stack, \texttt{tf2}, Nav2, and MoveIt~2. These systems are integrated by application developers who select a capability module, wire its interfaces to neighbouring modules, and deploy the result, rather than by the authors of each module. The contract model is designed for exactly this integration boundary, where the failures that motivate this paper occur.

The contract model is robot-agnostic at the interface level. It does not describe a specific kinematic chain, sensor suite, or control law; it describes the interface a module exposes to the rest of the system along six dimensions (signature, behavior, resources, permissions, recovery, and version). Two robots with different mechanics but the same interface vocabulary, the same typed topics and services, TF frames, QoS settings, resource ownership, and versioned packages, are governed by the same contract reasoning. A contract written for one embodiment transfers to another when the target shares that vocabulary. We argue this transfer rather than demonstrate it, since every case validated in this paper is a ROS system. When the target uses a different communication substrate or a different frame and resource convention, the contract must be re-expressed in the new vocabulary before it applies, even though the checking procedures of Section~\ref{sec:compatibility} are unchanged.

This scope is deliberately bounded. ECM Contracts validate the interface assumptions that govern whether modules can be installed, composed, and upgraded together; they are not a behavioral verification method and do not certify that an individual module computes a correct result. The class above (ROS~1 / ROS~2 packages integrated into mobile bases, manipulators, and service robots) is the class on which the model is validated in this paper, and it is also the class in which the documented integration failures we study were reported. We return to the conditions under which the model extends beyond this class, and to the dimensions whose evidence is strongest within it, in Sections~\ref{sec:evaluation} and~\ref{sec:discussion}.
\subsection{Contract Model}
\label{sec:contract}

We define an ECM Contract as a structured specification that accompanies every Embodied Capability Module and encodes the assumptions, requirements, and guarantees necessary for safe installation, composition, invocation, and upgrade. Our contract model draws on the tradition of assume-guarantee reasoning in contract-based design for cyber-physical systems~\citep{sangiovanni2012,nuzzo2014,benveniste2018}, extending it with dimensions specific to embodied execution. Formally, an ECM Contract is a six-tuple:
\begin{equation}
    \ecmcontract(e) = (\mathit{Sig}_e,\ \mathit{Beh}_e,\ \mathit{Res}_e,\ \mathit{Perm}_e,\ \mathit{Rec}_e,\ \mathit{Ver}_e)
\end{equation}
where each dimension captures a distinct aspect of embodied execution. An ECM is considered \emph{contract-complete} if all six dimensions are explicitly declared or inherited through certified defaults. Table~\ref{tab:contract_overview} summarizes the six dimensions at a glance: the question each one answers, the kinds of fields it carries, and the classes of failure it guards against. Table~\ref{tab:contract_schema} at the end of this section gives the concrete field-level schema with representative values. The remainder of this section walks through each dimension, explaining why conventional interface descriptions are insufficient and what an ECM contract adds.

\begin{table}[htbp]
\centering
\caption{The six dimensions of an ECM Contract at a glance. Each dimension answers a distinct validity question and guards against a distinct class of runtime failure. Field-level schema is given in Table~\ref{tab:contract_schema}.}
\label{tab:contract_overview}
\small
\begin{tabularx}{\textwidth}{lXXX}
\toprule
\textbf{Dim.} & \textbf{Question answered} & \textbf{Representative fields} & \textbf{Failure class guarded} \\
\midrule
$\mathit{Sig}$ &
  What does this module compute, in what representation? &
  I/O schemas with units and coord. frames; state objects; timing; invocation mode &
  type/unit/frame mismatches; missing state exposure \\
\addlinespace
$\mathit{Beh}$ &
  Under what conditions does its behavior match its specification? &
  preconditions, postconditions, invariants, semantic assumptions, completion/handoff semantics &
  violated assumptions; mismatched handoff states \\
\addlinespace
$\mathit{Res}$ &
  What physical and computational resources does it need? &
  sensors, actuators, compute, timing, communication, exclusive locks &
  resource contention; timing/bandwidth starvation \\
\addlinespace
$\mathit{Perm}$ &
  What is it authorized to do, and what must governance grant? &
  physical, data, network, and operational permissions; audit obligations &
  unauthorised motion / access; governance gaps \\
\addlinespace
$\mathit{Rec}$ &
  How does it fail, and what can the system rely on when it does? &
  failure taxonomy, rollback target, retry policy, safe-stop, escalation, composition recovery contract &
  unhandled faults; incompatible recovery across modules \\
\addlinespace
$\mathit{Ver}$ &
  How does it relate to its predecessor and dependencies? &
  semver, compatibility class, dependency ranges, deprecation metadata, policy/resource change markers &
  silent breaking upgrades; unsafe substitutions \\
\bottomrule
\end{tabularx}
\end{table}

\subsubsection{Functional Signature ($\mathit{Sig}$)}

The functional signature describes the module's computational interface: what data it consumes and produces, in what types, units, and reference frames. Unlike a conventional function signature, an ECM signature additionally specifies input/output schemas with physical units (e.g., meters, radians, Newtons) and coordinate-frame identifiers (e.g., \texttt{map\_frame}, \texttt{base\_link}, \texttt{end\_effector}), observable state objects for monitoring and governance inspection, temporal properties (execution duration, timeout bounds, real-time constraints), and invocation mode (synchronous, asynchronous, or event-triggered). Our functional signature extends the interface description paradigm established by ROS service definitions~\citep{quigley2009,macenski2022} with richer semantic annotations. The signature dimension answers: \emph{What does this module compute, in what representation?}

\subsubsection{Behavioral Assumptions ($\mathit{Beh}$)}

The behavioral dimension encodes the semantic conditions under which the module operates correctly: preconditions that must hold before activation (e.g., ``gripper is open,'' ``robot is within 1m of target''), postconditions guaranteed upon successful completion (e.g., ``object is grasped''), invariants that must hold throughout execution (e.g., ``end-effector stays within workspace bounds''), implicit semantic assumptions about the world model or physics (e.g., ``objects are rigid,'' ``surface is flat''), and completion/handoff semantics (``held,'' ``stabilized,'' ``transfer-ready,'' or merely ``nominal success''). The formalization of behavioral pre/postconditions follows the Design-by-Contract tradition~\citep{meyer1992} and has analogues in formal specification languages for robotics~\citep{miyazawa2019,kressgazit2018}. Behavioral assumptions are often left implicit in embodied interfaces even though they govern correct operation. They answer: \emph{Under what conditions does this module's behavior match its specification?}

\subsubsection{Resource Requirements ($\mathit{Res}$)}

The resource dimension specifies the physical and computational resources a module requires, spanning sensors (types, models, mounting, data rates), actuators (DoF, force/torque limits, control modes), compute (CPU/GPU/memory, inference-time constraints), timing (control-loop frequency, maximum latency, synchronization), communication (bandwidth, IPC channels, shared memory), and exclusive resource locks (e.g., gripper, navigation controller). Compared with conventional deployment descriptors, $\mathit{Res}$ treats physical devices and real-time budgets as first-class interface concerns rather than runtime afterthoughts.

\subsubsection{Permission Boundaries ($\mathit{Perm}$)}

The permission dimension declares the access rights the module requires from the runtime's governance framework. Conceptually it descends from capability-based protection systems~\citep{levy1984capability}, in which an access right is an unforgeable token granted to a subject; the embodied setting extends this with physical and operational rights that have no analogue in conventional capability systems. It covers physical permissions (zones the robot may enter, proximity limits to humans, force limits), data permissions (access to sensor streams, maps, object databases, or user data), network permissions (cloud services, model endpoints, external APIs), operational permissions (irreversible actions, high-force operations, safety-critical maneuvers), and audit requirements (mandatory logging, traceability, review obligations). Permission boundaries answer: \emph{What is this module authorized to do, and what must the governance layer grant before activation?}

\subsubsection{Recovery Semantics ($\mathit{Rec}$)}

The recovery dimension defines the module's failure-handling contract: an enumerated failure taxonomy with severity classifications (transient, degraded, critical, fatal); a rollback specification indicating which transitions can be reversed and which safe state the system returns to; a retry policy (conditions, maximum attempts, parameter changes between retries); a safe-stop action for unrecoverable failure; escalation conditions that hand control to a human operator, a supervisory module, or an emergency stop; and a composition recovery contract describing how the module's recovery interacts with upstream and downstream modules. Formal specification and verification of autonomous robotic systems, including recovery behaviors, has been surveyed by Luckcuck et al.~\citep{luckcuck2019}; our approach complements such methods by encoding recovery expectations declaratively within the module interface. Recovery semantics answer: \emph{How does this module fail, and what can the system rely on when it does?}

\subsubsection{Version Compatibility ($\mathit{Ver}$)}

The version dimension carries the metadata needed for lifecycle management: a semantic version identifier following the embodied semver scheme (Section~\ref{sec:release}); a compatibility classification relative to the predecessor (fully compatible, resource-sensitive, policy-sensitive, recovery-sensitive, or breaking); dependency constraints expressed as version ranges over other ECMs, runtime services, or platform capabilities; deprecation metadata (status, end-of-support date, migration pointers); and explicit policy/resource change markers that flag whether permissions or resources have shifted relative to the previous release.

\subsubsection{Complete Contract Schema}

Table~\ref{tab:contract_schema} presents the concrete, field-level schema that implementations follow, with representative values drawn from the grasp ECM used throughout the paper.

\begin{table}[htbp]
\centering
\caption{ECM Contract Schema: Six Dimensions of Embodied Capability Interfaces. Column abbreviations: Dim.\ = Dimension; Req.\ = Required (Y = required field, N = optional).}
\label{tab:contract_schema}
\small
\begin{tabularx}{\textwidth}{llllX}
\toprule
\textbf{Dim.} & \textbf{Field} & \textbf{Type} & \textbf{Req.} & \textbf{Example} \\
\midrule
\multirow{5}{*}{$\mathit{Sig}$}
 & \texttt{module\_id} & string & Y & \texttt{ecm.grasp.basic} \\
 & \texttt{input\_schema} & JSON Schema + units & Y & \texttt{pose: \{x,y,z\} m, map\_frame} \\
 & \texttt{output\_schema} & JSON Schema + units & Y & \texttt{grasp\_state: \{success, force\_N\}} \\
 & \texttt{coord\_frame} & frame\_id & Y & \texttt{base\_link} \\
 & \texttt{timeout\_ms} & uint32 & Y & \texttt{5000} \\
\midrule
\multirow{4}{*}{$\mathit{Beh}$}
 & \texttt{preconditions} & predicate list & Y & \texttt{gripper.is\_open, dist < 1m} \\
 & \texttt{postconditions} & predicate list & Y & \texttt{object.is\_grasped} \\
 & \texttt{invariants} & predicate list & N & \texttt{ee in workspace\_bounds} \\
 & \texttt{handoff\_semantics} & enum & Y & \texttt{persistent\_hold} \\
\midrule
\multirow{4}{*}{$\mathit{Res}$}
 & \texttt{sensors} & sensor spec list & Y & \texttt{wrist\_rgbd@30Hz, ft\_sensor} \\
 & \texttt{actuators} & actuator spec list & Y & \texttt{6dof\_arm, parallel\_gripper} \\
 & \texttt{compute} & resource spec & Y & \texttt{GPU: 4GB, CPU: 2 cores} \\
 & \texttt{exclusive\_locks} & resource list & N & \texttt{gripper} \\
\midrule
\multirow{3}{*}{$\mathit{Perm}$}
 & \texttt{physical\_perms} & permission list & Y & \texttt{human\_prox: 0.5m} \\
 & \texttt{data\_perms} & permission list & N & \texttt{read: object\_db} \\
 & \texttt{audit\_req} & enum & N & \texttt{standard} \\
\midrule
\multirow{4}{*}{$\mathit{Rec}$}
 & \texttt{failure\_modes} & failure + severity & Y & \texttt{slip: transient, collision: critical} \\
 & \texttt{rollback\_state} & state id & Y & \texttt{pre\_grasp\_pose} \\
 & \texttt{retry\_policy} & policy spec & Y & \texttt{max: 3, strategy: new\_pose} \\
 & \texttt{escalation} & condition list & Y & \texttt{3 failures $\to$ human\_takeover} \\
\midrule
\multirow{4}{*}{$\mathit{Ver}$}
 & \texttt{version} & semver & Y & \texttt{2.1.0} \\
 & \texttt{compat\_level} & enum & Y & \texttt{resource\_sensitive} \\
 & \texttt{dependencies} & version ranges & Y & \texttt{arm\_driver >= 3.0} \\
 & \texttt{policy\_change} & bool & Y & \texttt{false} \\
\bottomrule
\end{tabularx}
\end{table}

%

\emph{ECM Contracts are not merely type signatures; they are execution-validity descriptors for embodied capabilities.} Each dimension targets a concern that is absent from conventional interface description languages: type, unit, and frame mismatches~\citep{kate2021physframe,schneider2023composable}, behavioral and configuration assumptions~\citep{canelas2024misconfig}, and resource requirements that surface only at runtime~\citep{timperley2024robust}, together with permission and recovery requirements grounded in safety-engineering mechanisms rather than in documented composition bugs (Section~\ref{sec:contract_structure}). The six dimensions are not, however, equally load-bearing, and we do not claim that they are. Our evaluation (Sections~\ref{sec:prototype}--\ref{sec:evaluation}) measures how often each dimension predicts a real third-party integration failure, and the structure of the model follows from that evidence rather than from design symmetry. The rest of this section makes that structure explicit.

\subsubsection{Evidence-Based Structure of the Contract Model}
\label{sec:contract_structure}

A six-dimensional interface model invites a fair objection: that the dimensions are postulated for completeness rather than earned by evidence. We address this objection directly. We validate each dimension against documented integration failures on two substrates: the ROBUST corpus of 221 documented ROS bugs~\citep{timperley2024robust}, and a corpus of 16 documented ROS~2 integration bugs that we screened from public issue trackers and developer reports, keeping only cases backed by a specific defect report (the screening criteria and excluded candidates are given in Section~\ref{sec:evaluation}). Each scored case in both substrates carries a public source reference (Appendix~\ref{app:ledger}). Contracts are reconstructed from each module's published interface, with the checker frozen and hash-pinned before reconstruction, so the dimensions cannot be tuned to the bugs they are scored against (the full protocol is given in Section~\ref{sec:evaluation}). A separate blind reconstruction study (Section~\ref{sec:eval_blind}) tests whether the scored conflicts are derivable from the interface alone rather than from knowledge of the documented failure. The result is that the six dimensions separate into three tiers, and we report the tier of each one rather than presenting all six as equally validated.

\noindent\textbf{Validated core: $\mathit{Res}$ and named-change $\mathit{Ver}$.} Two dimensions predict real failures that no schema-only, schema-plus-QoS, or semantic-versioning baseline catches. Resource requirements ($\mathit{Res}$) predict 2 of 4 resource-contention failures on the ROS~2 corpus where every baseline scores zero (for example, multiple writers claiming \texttt{/cmd\_vel} and conflicting \texttt{ros2\_control} command-interface claims); these are detectable from the contract and from nothing in the schema, and the resource conflict is the clearest baseline-missed win. Version compatibility ($\mathit{Ver}$) predicts 6 of 18 version-attributed ROBUST failures and 5 of 7 on the ROS~2 corpus, against at most one for a semantic-versioning baseline that sees only declared major bumps. We restrict the version dimension to documented named-entity interface changes (a renamed, removed, or relocated topic, service, parameter, message field, class, or API symbol that a consumer still depends on), after a hash-versus-semver heuristic in the original checker proved unreliable (Section~\ref{sec:eval_rq2}). The retained cases are real interface-affecting deltas behind unchanged schemas, including the \texttt{ros2\_control} chainable-interface API change, a \texttt{mavros} \texttt{gps}-to-\texttt{global\_position} topic rename with a removed service, and MoveIt parameter-namespace moves. Functional signature ($\mathit{Sig}$) covers real type, unit, and frame mismatches (4 of 14 on ROBUST, 0 of 5 on ROS~2 after an independent check removed a constructed archetype) and is partly co-detected by a schema baseline, which is expected because it is the dimension closest to a conventional interface; we therefore report it alongside the core rather than as a baseline-missed win. $\mathit{Res}$ and named-change $\mathit{Ver}$ are the dimensions that earn the model its margin: they are where contract-level checking predicts failures that interface-level checking cannot.

\noindent\textbf{Scoped dimension: $\mathit{Beh}$.} Behavioral assumptions earn their place only for sub-properties that the runtime does not already enforce, and we scope the dimension accordingly. On the ROS~2 corpus, $\mathit{Beh}$ predicts 3 of 6 behavioral failures, and a strong schema-plus-QoS baseline predicts the same 3 cases, because the data-distribution-service (DDS) middleware underneath ROS~2 already enforces quality-of-service (QoS) reliability and durability compatibility at subscription time. The marginal contribution of $\mathit{Beh}$ beyond what the middleware enforces for free is zero on this corpus: every behavioral bug the contract predicts is a QoS property the middleware already checks. On the ROBUST substrate $\mathit{Beh}$ predicts zero failures, for a structural reason: ROBUST is ROS~1, which has no QoS profiles, so the reconstructed contracts carry no QoS field for the checker to test. We draw the corresponding conclusion: QoS-style behavioral compatibility is largely subsumed by the ROS~2 transport and should not be claimed as a contribution of the contract layer, whereas non-QoS behavioral assumptions (rates, ordering, world-model preconditions) remain the contract's responsibility because no middleware enforces them. $\mathit{Beh}$ is thus retained but deliberately narrowed, and its QoS sub-checks are positioned as a thin redundancy over the transport rather than as novel coverage.

\noindent\textbf{Forward-looking dimensions: $\mathit{Perm}$ and $\mathit{Rec}$.} Two dimensions are included on the strength of safety-engineering mechanisms rather than documented composition bugs, and we label them as such. Permission boundaries ($\mathit{Perm}$) had no documented multi-module composition failures in either corpus; the dimension is grounded instead in established access-control and safety mechanisms, namely SROS2 policy enforcement for ROS~2~\citep{mayoralvilches2022sros2}, the collaborative-operation and safety requirements of ISO~10218 for industrial robots~\citep{iso10218}, and the functional-safety lifecycle of IEC~61508~\citep{iec61508}. Recovery semantics ($\mathit{Rec}$) is runtime-leaning: it predicts zero failures under a static checker on both corpora (0 of 2 on the ROS~2 corpus), because recovery incompatibilities manifest during execution rather than in declared interfaces, and the same standards motivate specifying them in advance. We therefore present $\mathit{Perm}$ and $\mathit{Rec}$ as forward-looking dimensions whose value is argued from safety-standard mechanisms and whose empirical confirmation needs runtime evidence, which a static checker cannot supply. We do not count them toward the model's measured predictive margin.

We report the tiers this way so that the evidence for each part of the model is visible rather than pooled. An interface model for a safety-relevant domain should declare which of its parts are confirmed by evidence, which are scoped by what the runtime already provides, and which are anticipatory. Two properties make the tiers credible. First, the checker accepts every matched-good composition in a negative-control set, for a false-positive rate of zero on that control set (Section~\ref{sec:evaluation}); the model does not reach its margin by flagging everything. Second, each scored claim above is tied to a documented third-party failure, and the forward-looking dimensions are tied to a named safety standard rather than presented as measured. Table~\ref{tab:dim_evidence} records the tier, the supporting evidence, and the verdict for each of the six dimensions.

\begin{table}[!ht]
\centering
\caption{Evidence-based status of the six ECM contract dimensions. Recall is the fraction of in-scope documented integration failures that the dimension correctly predicts and attributes; ``ECM-unique'' marks dimensions on which every schema-only, schema-plus-QoS, and semantic-versioning baseline scores zero. Column abbreviations: Dim.\ = Dimension; ROBUST = ROS~1 corpus of \citet{timperley2024robust}; Corpus = ROS~2 documented-bug corpus.}
\label{tab:dim_evidence}
\small
\begin{tabularx}{\linewidth}{@{}l l l X@{}}
\toprule
\textbf{Dim.} & \textbf{Tier} & \textbf{Recall (ROBUST / Corpus)} & \textbf{Evidence and verdict} \\
\midrule
$\mathit{Res}$ & Core &
  -- / 2/4 (ECM-unique) &
  Resource contention (multi-writer \texttt{/cmd\_vel}, \texttt{ros2\_control} claims) predicted where every baseline scores zero; the cleanest baseline-missed win. \\
\addlinespace
$\mathit{Ver}$ & Core &
  6/18 / 5/7 &
  Restricted to documented named-entity interface changes behind unchanged schemas (service renames, parameter moves, API renames); beats a semantic-versioning baseline. \\
\addlinespace
$\mathit{Sig}$ & Core (co-detected) &
  4/14 / 0/5 &
  Real type, unit, and frame mismatches; partly co-detected by a schema baseline, as expected for the most interface-like dimension. \\
\addlinespace
$\mathit{Beh}$ & Scoped &
  0/8 / 3/6 (marginal 0) &
  QoS sub-checks are co-detected by a schema-plus-QoS baseline because DDS enforces QoS compatibility; scoped to non-QoS properties (rate, ordering). \\
\addlinespace
$\mathit{Perm}$ & Forward &
  -- / -- (no case) &
  No documented composition failures; grounded in SROS2~\citep{mayoralvilches2022sros2}, ISO~10218~\citep{iso10218}, and IEC~61508~\citep{iec61508}. \\
\addlinespace
$\mathit{Rec}$ & Forward &
  0 / 0/2 &
  Runtime-leaning; recovery incompatibilities appear in execution, not in declared interfaces; a static checker cannot confirm them. \\
\bottomrule
\end{tabularx}
\end{table}

The contract model therefore has a clear centre of gravity. Its predictive value over interface-level checking comes from $\mathit{Res}$ and the named-change part of $\mathit{Ver}$, both of which predict failures that no baseline catches, with $\mathit{Sig}$ contributing real but partly schema-co-detected coverage. $\mathit{Beh}$ contributes a thin non-QoS margin over what the ROS~2 transport already enforces. $\mathit{Perm}$ and $\mathit{Rec}$ are anticipatory dimensions justified by safety-standard mechanisms and intended for runtime checking. We keep all six because a contract is also a place to declare expectations the runtime should later enforce, but we report each dimension at the evidential tier the data support, and our headline claims rest on the validated core.
\subsection{Compatibility for Installation, Composition, and Upgrade}
\label{sec:compatibility}

\subsubsection{Overview}

A stable embodied capability ecosystem requires a principled way to determine whether a capability module can be installed, invoked, composed, or upgraded without violating the assumptions of the target runtime, dependent planners, active policies, or downstream modules. For embodied execution, conventional type-level or schema-level checks are insufficient because module validity depends on more than data formats: it also depends on behavioral assumptions, physical resource availability, permission boundaries, and failure-handling semantics.

Component-based robotics middleware such as OROCOS~\citep{bruyninckx2003} and SkiROS~\citep{rovida2017} provide composition mechanisms with varying levels of interface checking~\citep{elkady2012}; our framework generalizes these to six contract dimensions. We introduce a compatibility framework built on ECM Contracts that evaluates compatibility across the six contract dimensions and applies them to four lifecycle operations: (1)~\emph{installation compatibility}, (2)~\emph{invocation compatibility}, (3)~\emph{composition compatibility}, and (4)~\emph{upgrade compatibility}. Figure~\ref{fig:compat_overview} summarises the framework end-to-end: the six-dimensional contract is the single information source that all four lifecycle checks consult, parameterised by the target runtime, policy environment, and dependent module set, with each check yielding one of four structured outcomes. Sections~\ref{sec:compat_dim}--\ref{sec:compat_outcome} elaborate each component in turn, and every subsequent subsection refers back to this figure.

\begin{figure}[!ht]
\centering
\resizebox{\linewidth}{!}{%
\begin{tikzpicture}[
  font=\footnotesize,
  hub/.style    ={draw=tabBlue, very thick, rounded corners=3pt,
                  fill=white, minimum width=14cm, minimum height=0.8cm,
                  align=center, font=\small},
  side/.style   ={draw=tabGray!80!black, thick, rounded corners=2pt,
                  fill=white, minimum width=14cm, minimum height=0.8cm,
                  align=center},
  op/.style     ={draw=tabOrange, very thick, rounded corners=2pt,
                  fill=white, text width=2.5cm, minimum width=3.2cm, minimum height=1.2cm,
                  align=center, font=\small},
  outbox/.style ={draw=tabGreen, very thick, rounded corners=2pt,
                  fill=white, minimum width=14cm, minimum height=0.8cm,
                  align=center},
  arr/.style    ={-{Latex[length=2mm]}, thick},
]
\node[hub] (hub) at (0, 0)
  {\textbf{Six-dimensional ECM Contract}\,---\,$\mathcal{C}(e) = (\mathit{Sig},\,\mathit{Beh},\,\mathit{Res},\,\mathit{Perm},\,\mathit{Rec},\,\mathit{Ver})$};
\node[side, below=0.45cm of hub] (rpd)
  {parameterised by\,\,Runtime $R$,\,\,Policy $P$,\,\,Dependent set $D$};

\node[op] (inst) at (-5.4, -3.3)
  {\textbf{Installation}\\\scriptsize(\S\ref{sec:compat_install})\\[1pt]{\scriptsize\itshape static / pre-deploy}};
\node[op] (inv)  at (-1.8, -3.3)
  {\textbf{Invocation}\\\scriptsize(\S\ref{sec:compat_invoke})\\[1pt]{\scriptsize\itshape\bfseries runtime / live-state}};
\node[op] (comp) at ( 1.8, -3.3)
  {\textbf{Composition}\\\scriptsize(\S\ref{sec:compat_compose})\\[1pt]{\scriptsize\itshape static / pre-deploy}};
\node[op] (upg)  at ( 5.4, -3.3)
  {\textbf{Upgrade}\\\scriptsize(\S\ref{sec:compat_upgrade})\\[1pt]{\scriptsize\itshape static / pre-deploy}};
\foreach \n in {inst, inv, comp, upg} {
  \draw[arr] (rpd.south) -- (\n.north);
}

\node[outbox] (out) at (0, -4.95)
  {\textbf{Structured outcome:}~~Accept\,$|$\,Accept with conditions\,$|$\,Reject\,$|$\,Migration\,/\,review~~(\S\ref{sec:compat_outcome})};
\foreach \n in {inst, inv, comp, upg} {
  \draw[arr] (\n.south) -- (out.north);
}
\end{tikzpicture}%
}
\caption{Compatibility framework overview. The six-dimensional ECM contract is the single information source consulted at four lifecycle moments: installation, invocation, composition, and upgrade. Installation, composition, and upgrade are static gates that run before deployment; invocation is the single runtime gate, evaluated against live state at call time (tagged in each box). Each check is parameterised by the runtime $R$, policy $P$, and dependent module set $D$, and yields one of four structured outcomes. Sections~\ref{sec:compat_dim}--\ref{sec:compat_outcome} detail the dimension-to-operation mapping, the per-operation rules, and the outcome semantics in turn.}
\label{fig:compat_overview}
\end{figure}

\subsubsection{Compatibility Dimensions}\label{sec:compat_dim}

We treat compatibility as a multidimensional property. For an ECM $e$, let its contract be
\begin{equation}
    \ecmcontract(e) = (\mathit{Sig},\ \mathit{Beh},\ \mathit{Res},\ \mathit{Perm},\ \mathit{Rec},\ \mathit{Ver})
\end{equation}
as defined in Section~\ref{sec:contract}. Given a target runtime $R$, policy environment $P$, and optional dependent set $D$, the framework checks each of the six contract dimensions for consistency: signature schemas and frame alignment, behavioral pre/postcondition satisfaction, resource availability, permission authorization, recovery protocol compatibility, and version constraint satisfaction. Table~\ref{tab:compat_ops} summarizes which dimensions are checked for each operation.

\begin{table}[htbp]
\centering
\caption{Compatibility Operations and Checked Contract Dimensions.}
\label{tab:compat_ops}
\small
\begin{tabular*}{\linewidth}{@{\extracolsep{\fill}}lcccccc@{}}
\toprule
\textbf{Operation} & $\mathit{Sig}$ & $\mathit{Beh}$ & $\mathit{Res}$ & $\mathit{Perm}$ & $\mathit{Rec}$ & $\mathit{Ver}$ \\
\midrule
Installation & & & \checkmark & \checkmark & & \checkmark \\
Invocation & \checkmark & \checkmark & \checkmark & \checkmark & \checkmark & \\
Composition & \checkmark & \checkmark & \checkmark & \checkmark & \checkmark & \\
Upgrade & \checkmark & \checkmark & \checkmark & \checkmark & \checkmark & \checkmark \\
\bottomrule
\end{tabular*}
\end{table}

\subsubsection{Installation Compatibility}\label{sec:compat_install}

This is the leftmost lifecycle gate in Figure~\ref{fig:compat_overview}: the runtime decides whether $e$ may be admitted to the registry at all. An ECM $e$ is installation-compatible with runtime $R$ under policy $P$ if and only if:
\begin{enumerate}[leftmargin=*]
    \item All declared dependencies in $\mathit{Ver}_e$ are satisfiable in $R$.
    \item All required sensors, actuators, compute, and timing in $\mathit{Res}_e$ are supportable by $R$.
    \item All requested permissions in $\mathit{Perm}_e$ are allowed by $P$.
    \item Required contract fields are present (contract-completeness).
    \item No exclusive resource lock conflict exists with already-admitted modules.
\end{enumerate}
Formally:
\begin{multline}
    \textsc{InstallCompat}(e, R, P) = \textsc{DepSat}(e,R) \wedge \textsc{ResSat}(e,R) \\
    {}\wedge\; \textsc{PermSat}(e,P) \wedge \textsc{Complete}(e) \wedge \textsc{LockSafe}(e,R)
\end{multline}

This differs from ordinary software installation: admission depends on whether the target embodiment supports the module's operational assumptions, and installation has governance significance.

\subsubsection{Invocation Compatibility}\label{sec:compat_invoke}

This is the second gate in Figure~\ref{fig:compat_overview}, evaluated at call time rather than admission time: an installed module may still be unsafe to invoke under the current world state. Given planner state $s$, ECM $e$, runtime $R$, and policy $P$, invocation is compatible if:
\begin{enumerate}[leftmargin=*]
    \item The caller provides all required inputs with compatible schema and units.
    \item The preconditions in $\mathit{Beh}_e$ hold in state $s$.
    \item The resources required by $\mathit{Res}_e$ are currently allocatable.
    \item The requested permissions remain valid under $P$.
    \item The planner can accommodate declared completion and failure semantics.
\end{enumerate}

A module may be installed yet not currently safe to call. For instance, a navigation ECM may be present but not invocation-compatible if localization confidence has dropped below its declared precondition threshold.

\subsubsection{Composition Compatibility}\label{sec:compat_compose}

This is the third gate in Figure~\ref{fig:compat_overview}, where two or more contracts are reconciled across a multi-module pipeline. Given ECMs $e_i$ and $e_j$, write $e_i \prec e_j$ to denote that $e_i$ is composed before $e_j$, with $e_i$'s output supplied to $e_j$'s input. The directed composition $e_i \prec e_j$ is pairwise composition-compatible if and only if:

\begin{itemize}[leftmargin=*]
    \item \textbf{Type compatibility.} The output schema of $e_i$ is assignable to the input schema of $e_j$, including type, unit, and coordinate frame alignment. Where frames differ, an explicit frame transformation must be registered.
    \item \textbf{Behavioral compatibility.} The postconditions guaranteed by $e_i$ satisfy the preconditions required by $e_j$. The semantic assumptions of $e_j$ must be consistent with the world-state transformations performed by $e_i$.
    \item \textbf{Resource compatibility.} The combined resource requirements do not exceed platform capacity. Shared resources must not be double-allocated.
    \item \textbf{Governance compatibility.} The union of permissions required by all modules is grantable under current policy.
    \item \textbf{Recovery compatibility.} Failure-recovery strategies are non-conflicting: (a)~rollback states are reachable; (b)~retry behaviors do not violate invariants; (c)~escalation conditions are hierarchically consistent.
\end{itemize}

Formally, for pairwise composition:
\begin{multline}
    \textsc{ComposeCompat}(e_i, e_j, R, P) = \textsc{SigMatch}(e_i,e_j) \wedge \textsc{BehMatch}(e_i,e_j) \\
    {}\wedge\; \textsc{ResJointSat}(e_i,e_j,R) \wedge \textsc{PermChainSafe}(e_i,e_j,P) \wedge \textsc{RecMatch}(e_i,e_j)
\end{multline}

For a finite ordered chain $C=(e_1,e_2,\ldots,e_n)$ with $e_1 \prec e_2 \prec \cdots \prec e_n$, composition compatibility holds if each adjacent pair is pairwise compatible and global constraints remain satisfiable:
\begin{multline}
    \textsc{ChainCompat}(e_1,\ldots,e_n,R,P) = \bigwedge_{k=1}^{n-1} \textsc{ComposeCompat}(e_k,e_{k+1},R,P) \\
    {}\wedge\; \textsc{GlobalSafe}(e_1,\ldots,e_n,R,P)
\end{multline}

Some incompatibilities emerge only globally: no adjacent pair may conflict, yet the overall chain may exceed the available latency budget, violate cumulative audit requirements, or introduce lock contention through repeated exclusive resource use.

\subsubsection{Upgrade Compatibility}\label{sec:compat_upgrade}

This is the rightmost gate in Figure~\ref{fig:compat_overview} and the only one that consumes the dependent set $D$ explicitly: upgrades must remain compatible with all already-deployed consumers of the previous version. Given ECM versions $e^v$ and $e^{v'}$, we say $e^{v'}$ is upgrade-compatible with $e^v$ under runtime $R$, policy $P$, and dependent set $D$ if:
\begin{itemize}[leftmargin=*]
    \item \textbf{Backward compatibility.} Every invocation and composition contract that held for the previous version~$v$ must continue to hold for~$v'$.
    \item \textbf{Resource-sensitive compatibility.} If $v'$ requires additional resources, the upgrade requires resource availability verification.
    \item \textbf{Policy-sensitive compatibility.} If $v'$ requires new permissions, the upgrade requires governance re-approval.
    \item \textbf{Recovery-sensitive compatibility.} If $v'$ changes failure modes, rollback states, or retry strategies, all compositions must be re-validated.
    \item \textbf{Planner-sensitive compatibility.} If $v'$ alters behavioral assumptions, downstream planner models may be invalidated and require re-checking.
\end{itemize}

Formally:
\begin{multline}
    \textsc{UpgradeCompat}(e^v, e^{v'}, R, P, D) = \textsc{DepSat}(e^{v'},R) \wedge \textsc{DepPreserve}(e^v,e^{v'},D) \\
    {}\wedge\; \textsc{PermDeltaSafe}(e^v,e^{v'},P) \wedge \textsc{ResDeltaSafe}(e^v,e^{v'},R) \wedge \textsc{RecDeltaSafe}(e^v,e^{v'})
\end{multline}

An upgrade triggering any sensitive classification is not blocked but flagged for the appropriate gate check before deployment proceeds.

\subsubsection{Compatibility Outcomes}\label{sec:compat_outcome}

The bottom lane of Figure~\ref{fig:compat_overview} shows that all four lifecycle checks share a common outcome vocabulary. Rather than a single Boolean, the framework returns structured outcomes:
\begin{itemize}[leftmargin=*]
    \item \textbf{Accept:} all required checks pass.
    \item \textbf{Accept with conditions:} compatible under declared constraints.
    \item \textbf{Reject:} incompatibility detected, no safe deployment path.
    \item \textbf{Require migration or review:} compatibility uncertain or conditional.
\end{itemize}

\subsubsection{Algebraic Properties of the Compatibility Operators}\label{sec:compat_algebra}

We close the section by stating the algebraic properties on which the four operators rely. Following the assume--guarantee contract algebra of Benveniste et al.~\citep{benveniste2018}, each operator is a Boolean predicate over six contract-dimension sub-checks; these properties are simple consequences of that conjunctive structure but are worth recording explicitly so that downstream reasoning, registry implementation, and reviewer scrutiny can rely on them.

\noindent\textbf{Composition order-independence.} The chain check is independent of the order in which the resolver reconciles adjacent pairs. Since $\textsc{ChainCompat}(e_1,\ldots,e_n,R,P)$ is the conjunction of the pairwise $\textsc{ComposeCompat}$ checks and the global predicate $\textsc{GlobalSafe}$ (Section~\ref{sec:compat_compose}), and conjunction is associative and commutative, the chain verdict does not depend on the parenthesisation:
\[
\begin{aligned}
\textsc{ChainCompat}(e_1,\ldots,e_n,R,P) ={}& \textsc{GlobalSafe}(e_1,\ldots,e_n,R,P) \\
&{}\wedge \bigwedge_{k=1}^{n-1}\textsc{ComposeCompat}(e_k,e_{k+1},R,P),
\end{aligned}
\]
which is manifestly bracketing-invariant. We state the property at the level of the \emph{check} rather than as an operator that merges contracts: the prototype reconciles pairwise edges and does not materialise a single merged ECM, so we make no claim about a closed-form contract-composition operator $\textsc{Compose}(e_j,e_k)$ that returns a new contract. Defining such a merge (how to combine two modules' schemas, resource and permission sets, recovery strategies, and version chains into one contract, and prove it associative) is a natural extension we leave to future work; what the resolver relies on here is only the bracketing-invariance of the pairwise chain check, which the conjunctive structure guarantees.

\noindent\textbf{Upgrade compatibility is a partial order.} The relation $e^{v} \preceq e^{v'} \;\Leftrightarrow\; \textsc{UpgradeCompat}(e^v, e^{v'}, R, P, D)$ is reflexive ($e^v \preceq e^v$ trivially) and transitive (if $e^{v} \preceq e^{v'}$ and $e^{v'} \preceq e^{v''}$ then all five $\textsc{*DeltaSafe}$ predicates compose along the version chain). It is \emph{not} symmetric: an upgrade that adds a new permission or a new resource requirement is one-directional, since the prior version cannot satisfy a constraint declared only in the successor. This asymmetry is what makes \emph{embodied} semantic versioning (Section~\ref{sec:release}) more restrictive than conventional semver, which is implicitly symmetric on backward-compatible deltas.

\noindent\textbf{Soundness, not completeness.} The checker is designed to be \emph{sound} with respect to declared contracts: at the implementation level it accepts a composition only when every per-dimension sub-check passes, so an accepted composition is contract-compatible by its declared fields. We claim this as an implementation property of the rule set, not a proven theorem. It is deliberately \emph{not} complete, because a conservative multi-dimensional check can reject a composition that would in fact run. The negative-control result in Section~\ref{sec:evaluation} bounds this conservatism: the checker accepts every matched-good twin it was shown, so its rejections track declared incompatibilities rather than blanket caution. We therefore characterise the checker as \emph{conservative-sound}, a stronger property than ``best-effort'' but weaker than a decision procedure. Tightening it further would require either richer per-dimension predicates or runtime evidence injected into the static checker, both of which are deferred to future work.

\subsection{Release Discipline for Embodied Capability Ecosystems}
\label{sec:release}

A modular capability architecture does not automatically yield a stable ecosystem. Even when individual ECMs are contract-complete and pairwise composable, the ecosystem can become unstable if releases are published, upgraded, deprecated, or promoted without discipline: a new release may expand permission scope, alter recovery logic, increase hardware demands, or tighten timing assumptions while leaving interface schemas intact. We therefore extend the contract model with a release discipline that governs how ECM versions are classified, staged, and retired within a registry. This discipline rests on three elements: an \emph{embodied} semantic versioning scheme; a two-axis promotion lattice formed by compatibility classes and release channels; and an evolution mechanism that links deprecation, migration, and gate-keeping through the registry. Each element is grounded in the six contract dimensions of Section~\ref{sec:contract}, so release decisions reduce to checks on declared contract fields rather than ad hoc judgments.

\subsubsection{Embodied Semantic Versioning}

Traditional semantic versioning assumes compatibility is primarily determined by externally visible APIs. For ECMs, we extend versioning from interface shape to \emph{contract semantics}:

\begin{itemize}[leftmargin=*]
    \item \textbf{Patch release ($x.y.Z$):} Bug fixes that do not alter any contract dimension.
    \item \textbf{Minor release ($x.Y.0$):} Enhancements that extend capability while preserving existing contract guarantees. Resource requirements may decrease but not increase.
    \item \textbf{Major release ($X.0.0$):} Changes that modify one or more contract dimensions in a way that may invalidate existing compositions, invocations, or governance approvals.
\end{itemize}

Software package ecosystems~\citep{prestonwerner2013} provide the conceptual basis, but the extension to embodied capabilities introduces constraints absent from conventional dependency management: a change classified as ``minor'' under standard semver may constitute ``major'' in embodied semver if it alters resource requirements, permission needs, or recovery behavior, even when the functional signature remains unchanged.

\subsubsection{Compatibility Classes and Release Channels}

Each new release is assigned a \emph{compatibility class} that summarizes how its contract dimensions differ from the previous version and specifies the action the registry should take: promote, revalidate, require governance review, or require migration (Table~\ref{tab:compat_classes}). Compatibility classes capture \emph{what changed}; \emph{release channels} capture \emph{where a release is allowed to run}. Because embodied execution has physical consequences, channel progression is staged explicitly (Table~\ref{tab:channels}): a release enters the Sandbox channel with only partial contracts, advances through Beta and Stable as compatibility, policy, upgrade, and rollback evidence accumulates, and reaches the Certified channel only after strong policy review and fully traced rollback procedures. Together, the two tables define a two-axis promotion lattice: compatibility class determines what analysis a release requires, while channel determines which deployments may consume it.

\begin{table}[htbp]
\centering
\caption{Release Compatibility Classes.}
\label{tab:compat_classes}
\small
\begin{tabular*}{\linewidth}{@{\extracolsep{\fill}}lccccc>{\raggedright\arraybackslash}p{2.0cm}@{}}
\toprule
\textbf{Class} & $\mathit{Sig}$ & $\mathit{Beh}$ & $\mathit{Res}$ & $\mathit{Perm}$ & $\mathit{Rec}$ & \textbf{Action} \\
\midrule
Fully compatible & same & same & same & same & same & promote \\
Resource-sensitive & same & same & changed & same & same & revalidate \\
Policy-sensitive & compat. & same & opt. & broader & same & gov.\ review \\
Recovery-sensitive & same & success ok & opt. & same & changed & revalidate \\
Breaking & changed & changed & opt. & opt. & opt. & migration \\
\bottomrule
\end{tabular*}
\end{table}

\begin{table}[htbp]
\centering
\caption{Release Channels and Required Evidence.}
\label{tab:channels}
\small
\begin{tabular*}{\linewidth}{@{\extracolsep{\fill}}lccccc@{}}
\toprule
\textbf{Channel} & \textbf{Contract} & \textbf{Compat.} & \textbf{Policy} & \textbf{Upgrade} & \textbf{Rollback} \\
\midrule
Sandbox & partial & optional & optional & optional & optional \\
Beta & complete & limited & limited & limited & recommended \\
Stable & complete & verified & reviewed & tested & required \\
Certified & complete & verified & strong & tested & required + traced \\
\bottomrule
\end{tabular*}
\end{table}

\subsubsection{Evolution: Deprecation, Migration, and Release Gates}

Orderly ecosystem evolution requires that aging versions leave the system without breaking consumers. Each module version carries a deprecation flag for at least one release cycle before it may be removed from the registry, and every major release includes a machine-readable migration specification. Releasing parties are expected to provide adapter modules that translate old contracts into new ones where feasible, and the runtime supports simultaneous installation of two versions during migration periods so that dependent compositions can transition incrementally. Migration itself takes different forms depending on which contract dimension changed: schema migration transforms altered signature fields, units, or frames; behavioral adaptation bridges completion semantics; policy migration handles new permissions; resource migration reassigns execution to stronger embodiments; and recovery migration updates planner-level retry and rollback logic. The registry records these transitions by publishing, for each release, the full contract augmented with hardware tags, policy tags, certification status, environment tags, and a recovery profile summary, so that downstream tooling can evaluate upgrade safety mechanically rather than by reading release notes.

Promotion between channels is guarded by explicit \emph{release gates} (a contract completeness gate, a compatibility gate, a policy gate, a resource gate, a recovery gate, and a rollback gate), each derived from one or more of the six contract dimensions. A release is admitted to the next channel only when all applicable gates pass on the declared contract and on the evidence collected during the current channel, which makes channel promotion a mechanical, auditable process rather than a judgment call. The prototype in Section~\ref{sec:prototype} instantiates these gates as a single Release Gate Manager that consumes contract-class decisions, per-channel evidence, and policy profiles to produce an auditable promotion verdict.

\subsection{Prototype Design and Implementation}
\label{sec:prototype}

\subsubsection{Design Goals}

The prototype tests a central claim: ECM Contracts should be actionable system artifacts, not merely descriptive documentation. The implementation satisfies four goals: (1)~machine-checkable contracts, (2)~pre-deployment compatibility reasoning, (3)~release management connected to runtime governance, and (4)~lightweight integration with embodied runtime stacks.

\subsubsection{System Overview}
\label{subsec:system_overview}

Figure~\ref{fig:prototype_arch} gives the overall architecture of the reference prototype. Contract information flows along a single path from development to deployment: a developer packages an ECM together with its manifest; the registry stores the manifest and tracks release state; the resolver and compatibility checker decide, before deployment, \emph{what} can be selected and \emph{what} can be safely used together; the release gate manager governs promotion between channels; and the runtime validator re-checks, at execution time, the contract conditions that are only meaningful under live state. Two cross-cutting inputs, \emph{embodiment profiles} (available sensors, actuators, compute class, control rates) and \emph{policy profiles} (allowed motion scopes, human-contact rules, network access, release-channel restrictions), parameterize every deployment-time check so that the same ECM release can be evaluated against different contexts. The architecture reflects the core thesis: contract information should travel with an ECM throughout its lifecycle, rather than being discarded after development.

\begin{figure}[!ht]
\centering
\resizebox{\linewidth}{!}{%
\begin{tikzpicture}[
  font=\footnotesize,
  stage/.style   ={draw=tabBlue, very thick, rounded corners=2pt,
                   fill=white,
                   minimum width=2.15cm, minimum height=0.95cm, align=center},
  store/.style   ={draw=tabOrange, very thick, rounded corners=2pt,
                   fill=white,
                   minimum width=2.15cm, minimum height=0.95cm, align=center},
  profile/.style ={draw=tabGreen, very thick, dashed, rounded corners=2pt,
                   fill=white,
                   minimum width=2.15cm, minimum height=0.75cm, align=center},
  phase/.style   ={font=\scriptsize\itshape, tabGray!90!black},
  arr/.style     ={-{Latex[length=2mm]}, thick},
  farr/.style    ={-{Latex[length=2mm]}, thick, dashed, tabBlue!80!black},
  prarr/.style   ={-{Latex[length=2mm]}, thick, tabGreen!70!black},
]
\node[stage]               (manifest) at (0,0)    {ECM Manifest\\(YAML, 6 dims)};
\node[store]               (registry) at (2.6,0)  {Contract\\Registry\\[1pt]{\tiny SQLite\,+\,Git}};
\node[stage]               (resolver) at (5.2,0)  {Dependency\\Resolver};
\node[stage]               (checker)  at (7.8,0)  {Compatibility\\Checker\\[1pt]{\tiny Python, $\sim$1.2k LOC}};
\node[stage]               (gate)     at (10.4,0) {Release Gate\\Manager};
\node[stage]               (runtime)  at (13.0,0) {Runtime\\Validator\\[1pt]{\tiny ROS\,2 node / SROS2}};

\node[phase, above=0.08cm of manifest] {authoring};
\node[phase, above=0.08cm of registry] {publication};
\node[phase, above=0.08cm of resolver] {pre-deploy};
\node[phase, above=0.08cm of checker]  {pre-deploy};
\node[phase, above=0.08cm of gate]     {release};
\node[phase, above=0.08cm of runtime]  {execution};

\draw[arr] (manifest) -- (registry);
\draw[arr] (registry) -- (resolver);
\draw[arr] (resolver) -- (checker);
\draw[arr] (checker)  -- (gate);
\draw[arr] (gate)     -- (runtime);

\draw[farr] (runtime.north) -- ++(0,0.85)
            -| (registry.north)
            node[pos=0.25, above, font=\scriptsize]{promotion / audit evidence (QoS, rate, timeout telemetry)};

\node[profile] (embod)  at (5.2, -2.1) {Embodiment profile};
\node[profile] (policy) at (7.8, -2.1) {Policy profile};
\draw[prarr] (embod.north)  -- (resolver.south);
\draw[prarr] (policy.north) -- (checker.south);
\end{tikzpicture}%
}
\caption{Reference prototype architecture. Solid arrows trace the main flow: contract information travels from manifest authoring through the registry, resolver, and compatibility checker to the release gate manager, and then to the runtime validator at execution time. The dashed blue arrow closes the feedback loop, carrying audit evidence from the runtime back to the registry where it feeds subsequent promotion and deprecation decisions. Embodiment and policy profiles (green, dashed) parameterize pre-deployment compatibility checks and are also consulted by the runtime validator when evaluating live-state contract conditions.}
\label{fig:prototype_arch}
\end{figure}

\subsubsection{ECM Manifest and Registry}
\label{subsec:manifest_registry}

Each ECM is packaged with a machine-readable manifest that declares all six contract dimensions; Listing~\ref{lst:manifest} shows a simplified manifest for a navigation module. The Contract Registry stores these manifests alongside the corresponding code artifacts and version tags, together with release-channel state, validation evidence, and traceability information. Because the registry carries the manifest rather than only the binary, it can answer governance queries such as \emph{``which versions are admissible under a specific policy?''} and \emph{``which downstream ECMs may be affected by a proposed upgrade?''}

\begin{figure}[!ht]
\refstepcounter{figure}\label{lst:manifest}
\noindent\rule{\linewidth}{0.8pt}\par\vspace{1pt}
\noindent\textbf{Listing~\thefigure:}\ Simplified ECM manifest (\texttt{ecm.navigation.precise}, v1.3.0) declaring all six contract dimensions in YAML.\par\vspace{2pt}
\noindent\rule{\linewidth}{0.4pt}\par\vspace{3pt}
{\scriptsize
\begin{verbatim}
ecm:
  module_id: ecm.navigation.precise
  version: 1.3.0
  release_channel: beta
  compatibility_class: resource-sensitive
signature:
  input_schema: [{target_pose: Pose3D}, {tolerance: Float}]
  output_schema: [{nav_status: NavStatus}, {final_pose: Pose3D}]
  coordinate_frame: map
behavior:
  preconditions: [localization_confidence >= 0.90, base_ready]
  postconditions: [goal_reached, orientation_stable]
  completion_semantics: stable_arrival
resources:
  required_sensors: [lidar, imu]
  required_actuators: [mobile_base]
  control_frequency_hz: 20
  exclusive_resource_lock: [base_controller]
permissions:
  motion_permission: true
  restricted_zone_access: false
recovery:
  failure_modes: [timeout, path_blocked, localization_drop]
  retry_policy: retry_once
  rollback_state: last_safe_waypoint
  safe_stop_action: brake_and_hold
versioning:
  dependency_constraints: {runtime: ">=2.1,<3.0"}
  resource_change_marker: true
\end{verbatim}
}\vspace{-6pt}
\noindent\rule{\linewidth}{0.8pt}
\end{figure}

\subsubsection{Resolution and Compatibility Checking}
\label{subsec:resolver_checker}

Given a requested module family together with the active runtime, embodiment, and policy profiles, the \emph{dependency resolver} filters and ranks available releases: resolution answers \emph{what could be selected}. The \emph{compatibility checker} then answers \emph{what can be safely used together}: it implements the four compatibility types from Section~\ref{sec:compatibility} by performing pairwise contract comparison across all six dimensions, and for composition chains propagates contract deltas through the chain. For every detected incompatibility the checker reports the offending dimension, severity, and a suggested resolution. The engine is a standalone library and can be invoked by a registry service, a CI/CD pipeline, or a ROS~2 orchestration layer~\citep{macenski2022}, and integrates with runtime verification frameworks~\citep{caldas2024}.

\subsubsection{Release Gating and Runtime Validation}
\label{subsec:gate_runtime}

The \emph{release gate manager} enforces discipline over publication and promotion through a sequence of gate checks ($\text{Submit} \to \text{Validate} \to \text{Classify} \to \text{Promote}$). A beta-to-stable promotion, for example, requires contract completeness, successful composition checks on the target embodiments, no unresolved policy-sensitive deltas, and a rollback-ready deployment path. The \emph{runtime validator} complements these pre-deployment checks by validating contract conditions that are only meaningful under live state (current precondition satisfaction, lock availability, sensor availability, observed timeout paths), and records execution evidence that feeds back into release promotion, deprecation, and post-incident analysis (dashed arrows in Figure~\ref{fig:prototype_arch}).

\subsubsection{Implementation Scope}

The prototype is implemented in Python~3.10 with YAML-based manifest parsing and a deterministic rule-based checker; ECM manifests are authored as structured YAML documents covering all six contract dimensions (see supplementary materials for additional examples). The checker implements pairwise and chain-level composition checking, upgrade classification with contract-delta analysis, and dimension ablation support. Total implementation comprises approximately 1{,}200 lines of checking and experiment code. The prototype assumes manifests are truthfully declared and performs lightweight static reasoning over contract fields rather than full formal verification; runtime monitoring remains complementary.

\section{Evaluation}
\label{sec:evaluation}

An evaluation that labels its own ground truth and builds its own modules cannot show that contracts predict real failures. This section confronts that limitation directly. We set aside any author-synthetic benchmark and instead test whether ECM contract checking predicts integration failures that third parties documented independently, on modules built by third parties, with contracts reconstructed from each module's published interface and scored by a checker frozen before reconstruction. The numbers are moderate by design: a checker that genuinely predicts a documented class of integration bugs should recover a meaningful share of the in-scope cases while strong schema and version baselines recover almost none, and it should accept a composition once the mismatch is fixed rather than flag everything. A separate blind reconstruction study (Section~\ref{sec:eval_blind}) then tests the central residual threat, that the scored conflicts were encoded by the authors from knowledge of the bug rather than derived from the interface.

\subsection{Research Questions}
\label{sec:eval_rqs}

We organise the study around three questions. \textbf{RQ1 (Predictive validity)} asks whether contract checking predicts real, third-party-documented integration failures more accurately than schema-only, schema-plus-QoS, or semantic-version baselines. \textbf{RQ2 (Dimensional accounting)} asks which contract dimensions carry the predictive signal, and whether all six earn their place or some fail an adversarial falsification test against a strong baseline. \textbf{RQ3 (Over-rejection)} asks whether the checker accepts correct compositions or instead achieves recall by rejecting indiscriminately. RQ1 and RQ2 are answered by Path~B, a blind prediction study on two independent real-bug substrates. RQ3 is answered by a matched negative-control set. We then confirm three predicted failures in live execution (Path~A): one behavioral, one on the ECM-unique resource dimension, and one on the signature dimension.

\subsection{Path~B Protocol: Blind Prediction on Real Bugs}
\label{sec:eval_pathb_protocol}

\noindent\textbf{Reducing the circularity.} An author-synthetic evaluation couples four roles in the same hands: the modules, the failure labels, the contracts, and the checker are all authored by the same team. Path~B separates three of these so that no design decision can be tuned against the answer, and the blind reconstruction study of Section~\ref{sec:eval_blind} addresses the fourth, contract reconstruction, directly. \emph{Modules} are real third-party ROS packages (kobuki, mavros, Universal Robots, motoman, turtlebot, cob), not artefacts we wrote. \emph{Ground truth} is the set of integration failures those projects documented themselves, taken from the ROBUST bug corpus of Timperley et al.~\citep{timperley2024robust} and from public ROS\,2 issue records that we collected but did not author, not labels we assigned (the per-case sources are listed in Appendix~\ref{app:ledger}). \emph{Contracts} are reconstructed from each package's published interface (message and service definitions, parameters, declared QoS), with the failure record withheld during reconstruction; because the authors performed this reconstruction, Section~\ref{sec:eval_blind} tests with independent annotators whether the scored conflicts are derivable from the interface alone. The \emph{checker} is frozen: its source is content-addressed and the hash is pinned before any contract is reconstructed, so the rules cannot be edited to fit a case after seeing it. Finally, the \emph{gold dimension} for each bug is assigned from each corpus's own fault taxonomy through a frozen crosswalk, blind to which dimension ECM would fire. A prediction counts as correct only when the checker both flags the composition and attributes the failure to the gold dimension; merely rejecting for the wrong reason does not count.

\noindent\textbf{Selection funnel.} The ROBUST substrate illustrates the in-scope reasoning, which we report transparently rather than after filtering to a favourable subset. The corpus contains 221~documented bugs. The frozen crosswalk marks 92~as integration-relevant; the remaining bugs are build, single-module logic, or firmware faults that no composition checker should claim. Of the 92, blind raters judged 37~assessable as runtime composition or upgrade faults, excluding the remaining 55 (of the 92) that are build, logic, or firmware in nature. Two cases could not be grounded in any interface property and were dropped with none forced, leaving 35~interface-property-scoreable cases. Of these 35, 32~carry a crosswalk gold dimension and form the scored set; the three dropped at this last step are topic-name and wiring faults that depend on launch-file configuration rather than the module interface, and so fall outside what a contract over interfaces can predict. We state these out-of-scope categories plainly because they bound what the method can claim (Section~\ref{sec:eval_threats}).

\noindent\textbf{Two substrates.} The two corpora probe complementary failure regimes. The \emph{ROBUST} substrate (ROS\,1) exercises signature and version faults across robot drivers and middleware: real version deltas such as the mavros \texttt{gps}~$\to$~\texttt{global\_position} rename with a removed service and several MoveIt parameter-namespace moves, each a documented named-entity change to a topic, service, or parameter that a consumer still depends on. ROS\,1 has no QoS profiles, so contracts reconstructed from it carry no QoS field; the behavioral QoS sub-checks therefore cannot fire on this substrate by construction. The \emph{ROS\,2 corpus} (16~source-cited real ROS\,2 bugs) supplies the regime ROS\,1 lacks: QoS reliability and durability mismatches, resource-claim conflicts, and recovery semantics under the ROS\,2 middleware~\citep{macenski2022}. Reporting both substrates separately, rather than pooling them, is what keeps the per-dimension story accountable: each dimension is credited only on the substrate that can actually exercise it.

\noindent\textbf{Corpus screening.} The ROS\,2 corpus was screened for provenance before scoring. We began from a larger candidate pool and retained only cases backed by a specific third-party defect report: a GitHub issue, a developer question-and-answer thread, or a bug catalogued in an empirical study. We excluded 14 candidates that did not meet this bar. Two were flagged as unverified; six were modelled from a failure-category taxonomy~\citep{canelas2024misconfig} rather than a documented instance; two described a mechanism in a design document rather than a reported defect; and four resolved on inspection to usage questions or user-configuration errors rather than project defects. The screening was applied on provenance grounds alone, independent of checker outcome. We then de-duplicated by underlying defect: two retained cases (a resource view and a version view of the same \texttt{ros2\_control} chainable-interface change, issue~\#1400) describe one defect, so we merged them into a single scored case to avoid double-counting it across dimensions, which leaves 17 cases. A later independent reconstruction (Section~\ref{sec:eval_recon}) removed one further case, a constructed frame archetype that is not a specific documented instance, leaving 16 scored ROS\,2 cases. The 16 retained cases and the 15 excluded candidates, each with its source and exclusion reason, are listed in the replication package.

\noindent\textbf{Methods compared.} We compare ECM-full (all six dimensions) against three baselines that represent the checking that real ROS toolchains already perform. \emph{Schema-only} matches interface types from the message and service definitions, the level enforced by \texttt{rosidl}. \emph{Schema+QoS} adds the DDS QoS compatibility rules the ROS\,2 middleware enforces at subscription time, making it a strong, non-strawman baseline for the behavioral dimension. \emph{Semver-only} applies conventional semantic-version rules~\citep{prestonwerner2013} to declared version numbers. These baselines are chosen so that any margin ECM shows is a margin over checking that practitioners can run today, not over a deliberately weak comparator.

\subsection{RQ1: Predictive Validity}
\label{sec:eval_rq1}

Table~\ref{tab:pathb_recall} reports recall on both substrates, where recall is the fraction of scored bugs that a method both predicts and attributes to the correct dimension. ECM-full recovers 10~of~32 documented bugs on the ROBUST substrate (31\%) and 10~of~16 on the ROS\,2 corpus (62\%). Every baseline stays low: schema-only at 0--6\%, schema+QoS at 6--19\%, and semver-only at 0--19\%. The strongest baseline on either substrate (schema+QoS and semver-only at 19\% on ROS\,2) recovers at most a third of what ECM-full recovers there. The margin holds across the two independent substrates, which is the predictive evidence that an author-synthetic accuracy figure cannot provide. Across both substrates the advantage is statistically significant: an exact McNemar test of ECM against the strongest baseline (schema+QoS) gives $p=0.0078$ on ROBUST ($b{=}8, c{=}0$) and $p=0.016$ on ROS\,2 ($b{=}7, c{=}0$), both significant at the $0.05$ level, so ECM wins cases the baseline misses and never loses a case the baseline wins. Because ECM-full is a superset of the baseline checks, this test should be read as evidence that the added contract dimensions raise the flag rate on these documented bugs, not as a claim of independence from the failure narrative; the blind reconstruction study in Section~\ref{sec:eval_blind} addresses that independence separately. The Wilson 95\% lower bound for ECM-full (18\% on ROBUST, 39\% on ROS\,2) and the baseline intervals are reported in Table~\ref{tab:pathb_stats}.

\begin{table}[!ht]\centering
\caption{Recall with Wilson 95\% confidence intervals and exact McNemar tests of ECM-full against the strongest baseline (schema+QoS).}
\label{tab:pathb_stats}\small
\begin{tabularx}{\linewidth}{@{}lXX@{}}\toprule
\textbf{Method} & \textbf{ROBUST ($N{=}32$)} & \textbf{ROS\,2 ($N{=}16$)} \\\midrule
ECM-full & 10/32 [18,49]\% & 10/16 [39,82]\% \\
schema-only & 2/32 [2,20]\% & 0/16 [0,19]\% \\
schema+QoS & 2/32 [2,20]\% & 3/16 [7,43]\% \\
semver-only & 0/32 [0,11]\% & 3/16 [7,43]\% \\
\midrule McNemar vs schema+QoS & $b{=}8,c{=}0,\ p{=}0.0078$ & $b{=}7,c{=}0,\ p{=}0.016$ \\
\bottomrule\end{tabularx}\end{table}

\begin{table}[!ht]\centering
\caption{Recall on both substrates (a method predicts a bug when it emits a critical finding whose dimension is in the bug's gold set).}
\label{tab:pathb_recall}\small
\begin{tabularx}{\linewidth}{@{}lXX@{}}\toprule
\textbf{Method} & \textbf{ROBUST ($N{=}32$)} & \textbf{ROS\,2 ($N{=}16$)} \\\midrule
ECM-full (6-dim) & 10/32 (31\%) & 10/16 (62\%) \\
schema-only & 2/32 (6\%) & 0/16 (0\%) \\
schema+QoS & 2/32 (6\%) & 3/16 (19\%) \\
semver-only & 0/32 (0\%) & 3/16 (19\%) \\
\bottomrule\end{tabularx}\end{table}

\begin{figure}[!ht]
\centering
\includegraphics[width=\linewidth]{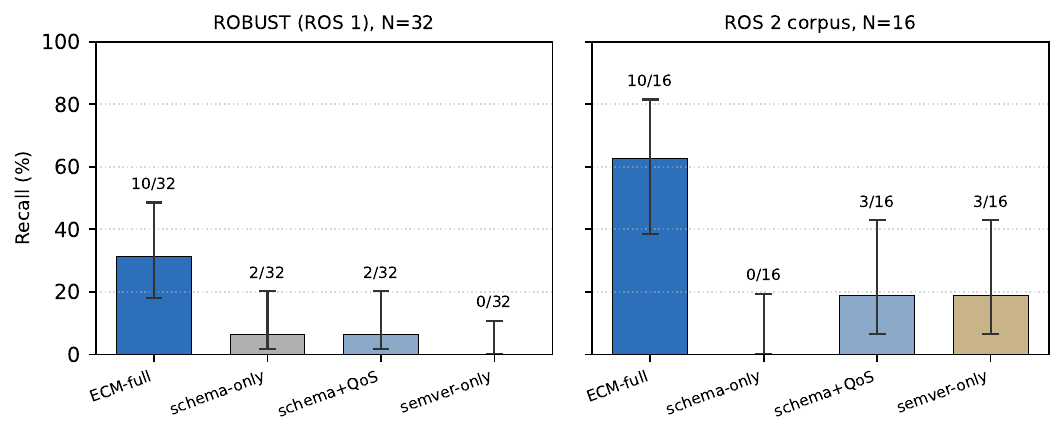}
\caption{Recall of ECM-full and the three baselines on both substrates, with 95\% Wilson confidence intervals; per-method counts are annotated above each bar. ECM-full's margin over the strongest baseline (schema+QoS and semver-only at 19\% on ROS\,2 and 6\% on ROBUST) is visible on both corpora, and the wide intervals reflect the small sample sizes discussed in Section~\ref{sec:eval_threats}. These are the same numbers reported in Tables~\ref{tab:pathb_recall} and~\ref{tab:pathb_stats}.}
\label{fig:results}
\end{figure}

\subsection{RQ2: Which Dimensions Earn Their Place}
\label{sec:eval_rq2}

Aggregate recall does not show whether all six dimensions contribute or whether a few carry the result. Table~\ref{tab:pathb_perdim} breaks recall down by dimension on the substrate that can exercise each one, and the reading is deliberately uneven.

\begin{table}[!ht]\centering
\caption{Per-dimension recall of ECM-full (multi-label denominators: a bug stressing two dimensions counts under each).}
\label{tab:pathb_perdim}\small
\begin{tabularx}{\linewidth}{@{}lXX@{}}\toprule
\textbf{Dimension} & \textbf{ROBUST} & \textbf{ROS\,2} \\\midrule
Signature (\textit{Sig}) & 4/14 & 0/5 \\
Behavioral (\textit{Beh}) & 0/8 & 3/6 \\
Resource (\textit{Res}) & --- & 2/4 \\
Permission (\textit{Perm}) & --- & --- \\
Recovery (\textit{Rec}) & --- & 0/2 \\
Version (\textit{Ver}) & 6/18 & 5/7 \\
\bottomrule\end{tabularx}\end{table}

\noindent\textbf{Version, restricted to named-entity changes.} The original checker flagged a version break whenever an interface hash changed without a major semantic-version bump. On audit this heuristic proved unreliable: it fired on modelled placeholder hashes, and a blind annotator given the same interface facts did not reproduce it (Section~\ref{sec:eval_blind}). We removed the heuristic and restrict the version dimension to documented named-entity changes: a renamed, removed, or relocated topic, service, parameter, message field, class, or API symbol that a consumer still depends on, which the checker detects through its \texttt{field\_break} rule. With this restriction the version dimension recovers 6~of~18 version-attributed bugs on ROBUST and 5~of~7 on ROS\,2, from real interface-bearing deltas (the \texttt{mavros} \texttt{gps}-to-\texttt{global\_position} rename with a removed service, MoveIt parameter-namespace moves, the \texttt{ros2\_control} chainable-interface API change). Semver-only recovers about one of these, because it can only see declared major-version bumps and not the named interface change behind them. The version-triggered cases without a named interface change (for example a GUI tool that crashes after an upgrade) are no longer counted, which is why version recall on ROBUST falls to 6 of 18; we prefer the smaller interface-self-evident count to the larger heuristic one.

\noindent\textbf{Resource is the cleanest baseline-missed win.} On ROS\,2, the resource dimension recovers 2~of~4 resource-attributed bugs, the multi-writer conflict on \texttt{/cmd\_vel} and a command-interface claim conflict in \texttt{ros2\_control}. Every baseline recovers zero of these, because resource contention is invisible to type and version checking. The fourth resource case is a dual-arm \texttt{ros2\_control} fault whose two arms in fact claim disjoint per-joint interfaces; the independent reconstruction of Section~\ref{sec:eval_recon} showed it to be an unscoped-implementation defect rather than an interface-level resource conflict, so the contract correctly does not predict it and we count it as a miss. The genuine resource conflict remains the clearest case of a dimension predicting failures that nothing in the existing toolchain catches, and both the blind study and the independent reconstruction reproduce it from the interface alone (Sections~\ref{sec:eval_blind} and~\ref{sec:eval_recon}).

\noindent\textbf{Signature is real but scoped.} The signature dimension catches genuine type, frame, and unit mismatches but recovers only a moderate share (4/14 on ROBUST, 0/5 on ROS\,2). The documented signature wins are the ROBUST coordinate-frame faults (the \texttt{mavros} ENU-versus-NED attitude and velocity conversions, independently re-derived in Section~\ref{sec:eval_recon}); on the ROS\,2 corpus the dimension scores zero of five after the independent reconstruction removed the one frame case that was a constructed archetype rather than a documented instance (Section~\ref{sec:eval_recon}). The remaining shortfall has a clear cause: many signature-class bugs in these corpora are topic-name and namespace wiring faults fixed in launch configuration, which a contract over module interfaces cannot predict. We report the ceiling rather than reclassify wiring faults as interface faults.

\noindent\textbf{Behavioral fails the falsification test.} The behavioral dimension is where an adversarial reading matters most, because its main mechanism is QoS compatibility, which the ROS\,2 middleware already enforces for free. On the 6~behavioral cases, ECM catches 3 and the schema+QoS baseline catches the same 3, so ECM has no margin over QoS-aware checking on this corpus. The conclusion this forces is that the behavioral dimension does not clear a keep-as-core bar on its own: its QoS sub-checks are co-detected by a baseline a practitioner can already run, and only its non-QoS sub-properties add value. On the ROBUST substrate the behavioral dimension recovers 0~of~8, exactly as construction predicts, since ROS\,1 carries no QoS field for a reconstructed contract to encode. We therefore scope the behavioral dimension to its non-QoS sub-properties and fold the QoS check into the schema layer.

\noindent\textbf{Recovery and permission are forward-looking.} The recovery dimension recovers 0~of~2 on ROS\,2 and contributes nothing on ROBUST. The cause is structural: recovery faults are runtime-leaning (they surface as rollback or retry behaviour during execution) and a static interface checker cannot observe them. We keep the recovery dimension as conditional and runtime-dependent rather than claim static predictive value it does not have. The permission dimension has no documented composition bug of this class in either corpus after screening. We therefore treat permission as forward-looking, grounded in the access-control mechanisms of SROS2~\citep{mayoralvilches2022sros2} and in the functional-safety requirements of ISO~10218 and IEC~61508~\citep{iso10218,iec61508} rather than in measured prediction.

Read together, the data support a core anchored on the resource dimension and the named-change part of the version dimension, both of which predict failures no baseline catches, with the signature dimension contributing real but partly schema-co-detected coverage. The behavioral dimension is scoped to its non-QoS sub-properties, and the recovery and permission dimensions are held as forward-looking. This is a reduced, evidence-backed model in place of an unsupported six-dimension claim, and it states each falsification trigger plainly.

\subsection{RQ3: The Checker Does Not Over-Reject}
\label{sec:eval_rq3}

High recall is worthless if it comes from rejecting everything, which is the central risk for any checker that rejects aggressively. We test this with matched negative controls on the ROS\,2 corpus. For each scored bug we construct a matched-good twin in which the documented mismatch is neutralised while the rest of the composition is held fixed, then ask whether the checker still rejects. On all 16~matched twins, ECM raises zero false positives, a precision of 100\% on the control set. The checker accepts the fixed compositions it is supposed to accept; it does not flag good and bad compositions alike. Because each twin is built by removing the exact field the checker tests, this result establishes that the checker is not blindly conservative (it goes quiet when the declared mismatch is removed). It removes the over-rejection explanation for the recall in Table~\ref{tab:pathb_recall}: the dimensions that fire on real bugs go quiet when the mismatch is removed.

\noindent We add a second, stronger control that does not blank any field. For six bugs with a documented fixing commit we reconstruct the \emph{real post-fix} interface (the consumer rebuilt against the renamed or removed entity, the planning request relabelled to the model frame) and re-run the frozen checker. It accepts five of the six: acceptance here comes from the consumer genuinely tracking the new interface, not from removing the field the checker tests. The sixth, the \texttt{ros2\_control} \texttt{on\_init} lifecycle change, is still flagged, because the \texttt{field\_break} rule is name-granular: it sees that a consumer references the changed \texttt{on\_init} symbol and cannot tell that the consumer re-implemented it against the new signature. We report this case as a real-version false positive and as evidence of the rule's granularity limit (Section~\ref{sec:eval_threats}). On real fixed versions, then, the checker discriminates real fixed from real broken on five of six cases, a stronger precision result than the field-removal twins alone.

\subsection{Are the Conflicts Interface-Derivable? A Blind Reconstruction Study}
\label{sec:eval_blind}

The central residual threat to Path~B is that the contract fields were reconstructed by the authors, who knew the documented failure, so a scored win could encode the answer rather than predict it. To test whether the contract-field reconstruction encodes the documented failure (a circularity risk) or is instead derivable from each module's public interface, we ran a blind reconstruction study. We built 15~interface-only briefs. Each brief gives only the two modules' declared interfaces (message types, coordinate frames, units, declared exclusive hardware and resource claims, QoS, and, for upgrades, the version tags and changed fields), with the failure narrative and the gold dimension removed. We mixed documented-bug compositions with negative controls (matched-good twins and unrelated pairs). Independent annotators, blind to the bug, judged from the interface facts alone whether each composition would fail and on which dimension.

The signature, resource, and behavioral-QoS conflicts were independently reproduced from the interface facts in all six such cases: a frame mismatch, two resource double-claims, and three QoS incompatibilities. The negative controls were correctly cleared. This indicates that these conflicts are interface-derivable rather than bug-informed: an annotator with only the declared interfaces, and no knowledge of the bug, reaches the same prediction the checker does. The version cases that the original checker had flagged only through a hash-versus-semver heuristic were \emph{not} reproduced by the blind annotators. This is what led us to restrict the version dimension to documented named-entity changes (Section~\ref{sec:eval_rq2}): the named-entity changes are visible as concrete renamed or removed interface symbols in the brief, whereas a bare hash difference without a named change is not interface-self-evident and the annotators did not treat it as predictive. The study does not remove the reliability threat entirely, because the authors still reconstructed the full corpus; a complete independent reconstruction of every contract by a second team remains the strongest open validation (Section~\ref{sec:eval_threats}).

\subsection{Path~A: Runtime Confirmation}
\label{sec:eval_patha}

Path~B shows that the checker predicts documented failures; Path~A confirms that the predicted failure is a real runtime event decided by the middleware, not an author label. We reproduce the behavioral QoS case in live ROS\,2 Humble execution. A publisher with a \texttt{BEST\_EFFORT} reliability profile is composed with a subscriber requiring \texttt{RELIABLE} delivery, the exact incompatibility the contract flags. Repeated five times, the incompatible composition delivers 0 of 80 messages on every run while the \texttt{rmw} layer logs that it is requesting an incompatible QoS and that no messages will be sent; the matched-compatible composition (\texttt{RELIABLE} publisher to \texttt{RELIABLE} subscriber) delivers 79 of 80 messages on every run. The contract's prediction therefore corresponds to a real delivery failure adjudicated by the DDS middleware, closing the loop from a static contract check to an observed runtime outcome. We note that this QoS rule is the same DDS compatibility rule the schema+QoS baseline already encodes, so this case confirms that the contract's behavioral sub-check matches middleware behavior rather than demonstrating coverage unique to ECM; the unique-coverage confirmation is the resource case below.

\noindent We confirm a second prediction, on the resource dimension that no baseline catches. Two independently launched nodes publish velocity commands to the single \texttt{/cmd\_vel} actuation topic of a TurtleBot3 in Gazebo, which is the multi-writer resource conflict the contract flags. Across five repetitions, a single writer commanding forward motion advances the robot 0.37--0.77\,m in four seconds, whereas with the second writer added the two contend for the single-owner actuation resource and forward progress collapses to 0.02--0.08\,m on every run, with a consistent spurious yaw (up to 1.5\,rad), so neither writer's intent is realised. This confirms that the resource conflict, the strongest ECM-unique prediction, is a real actuation failure rather than only a static flag. Both outcomes are stable under repetition: the QoS incompatibility is enforced identically by the middleware on every connection, and the resource-conflict contrast (single-writer progress versus the two-writer collapse) holds on every one of the five runs.

\noindent We confirm the resource conflict a second way, reproducing the exact mechanism the documented \texttt{ros2\_control} bugs report. Using the real \texttt{ros2\_control} controller manager (version~2.54.0) over a mock hardware interface, we load two \texttt{ForwardCommandController}s that each declare the \texttt{joint1/position} command interface. The first activates; when the second is activated, the resource manager rejects it at runtime with \texttt{Resource conflict for controller 'ctrlB'. Command interface 'joint1/position' is already claimed}, the verbatim failure reported in the \texttt{ros2\_control} issues the resource dimension predicts. The exclusive command-interface claim is declared in each controller's published configuration, so the contract flags the conflict statically before the controller manager rejects it dynamically.

\noindent We add a third confirmation on the signature dimension, which the static checker catches and which surfaces as a quantifiable geometry error at runtime. A producer declares a target point in the \texttt{base\_link} frame and a consumer expects it in the \texttt{camera} frame. With the camera extrinsic (a $0.5$\,m and $0.2$\,m offset and a $90^{\circ}$ yaw) published through the ROS\,2 \texttt{tf2} transform tree, the correctly transformed point lands at $(-0.20, -0.50, 0)$\,m in the camera frame, whereas a consumer that ignores the frame mismatch acts on the raw $(1.0, 0, 0)$\,m, an end-effector error of $1.30$\,m computed through the real \texttt{tf2} stack. The frame fields the contract compares are declared in each module's interface, so the contract flags the mismatch statically; the runtime confirms it would otherwise drive a grasp or handover to a target more than a metre off. Across the three confirmations the resource conflict (the validated core that no baseline catches), the signature frame mismatch, and the behavioral QoS case each correspond to a real runtime event adjudicated by the production middleware rather than to an author label.

\subsection{Targeted Replication on Additional Documented Cases}
\label{sec:eval_replication}

The screened ROS\,2 corpus is small on the two dimensions that carry the predictive signal, so we ran a targeted replication to test whether the resource and named-version mechanisms generalise beyond it. We searched the public issue trackers of MoveIt~2, Nav2, and \texttt{rclcpp} for documented integration failures of the signature (coordinate-frame) and version (named-entity) classes, independently refetched each candidate to confirm that it documents a real defect, and admitted only cases whose conflict is a property of the module interface. Of ten documented candidates, four met this bar after an independent reconstruction check (Section~\ref{sec:eval_recon}): one coordinate-frame mismatch (the MoveIt~2 PILZ planner rejecting a goal whose \texttt{frame\_id} differs from the model frame) and three named-entity changes (the Nav2 \texttt{progress\_checker\_plugin} parameter rename, the \texttt{Spin.action} \texttt{disable\_collision\_checks} field, and the \texttt{rclcpp} \texttt{async\_send\_request} return-type change). We excluded six with reasons: one consumer-side transform bug that is not an interface incompatibility, one namespace-wiring fault and one timing-dependent failure that fall outside what a static interface contract can predict, one ROS\,1-era case that would mix substrates, one duplicate, and one Nav2 goal-frame case that the independent reconstruction found to use matching default frames (so there is no interface-level mismatch unless a user sets a non-default frame). Scored by the same frozen checker, the contract predicts all four admitted cases from their declared interfaces alone and attributes each to its gold dimension (one signature, three version). Because these cases were found by searching for documented instances of the classes the contract targets, this is a test of the mechanism's breadth on independently sourced real cases, not an unbiased recall estimate; we keep it separate from the headline corpus for that reason. The four admitted cases, the six exclusions with their reasons, and the per-field interface evidence are listed in the replication package.

\subsection{Independent Reconstruction of the Scored Wins}
\label{sec:eval_recon}

The blind study of Section~\ref{sec:eval_blind} used interface briefs that the authors prepared. To remove that last dependency, we ran an independent reconstruction and extended it from the validated core to \emph{every} case the checker scores as a win on either substrate, since a circular encoding can only matter where the checker fires (a missed case cannot encode the answer). An automated agent, given only each module's public identity and instructed to record interface fields as a neutral author who is not told a bug exists, fetched the public artefacts (message, service, and action definitions, controller configurations, version diffs, QoS profiles, and coordinate-frame conventions) and re-derived the checker-relevant fields (the coordinate \texttt{frame}, the exclusive resource tokens, the \texttt{changed\_fields}, and the \texttt{consumer\_reads}) for each win. The agent was blind both to the documented failure and to the authors' own encoding. We then re-ran the frozen checker on the re-derived fields and compared the outcome to the authored encoding. This is an agent-based reconstruction, not a second human team, and we report it as such.

Of the twenty wins that remain after the corrections below, every one was independently corroborated, each citing the concrete public source: the \texttt{joint1/position} command-interface claim for the resource case; the ENU-versus-NED attitude and velocity conversions in \texttt{mavros}; the \texttt{gps}-to-\texttt{global\_position} topic rename, the \texttt{SetMode} \texttt{success}-to-\texttt{mode\_sent} field rename, and the removed \texttt{WaypointGOTO} service, each traced to the commit that changed it; the MoveIt parameter-namespace move; the \texttt{on\_init} lifecycle-signature change; and the QoS reliability mismatches confirmed against the published \texttt{SensorDataQoS} profiles. No retained win failed to reproduce. This is direct, outcome-independent evidence that the scored conflicts are reconstructible from the public interface rather than supplied by the authors.

Three cases did not survive the independent check, and all three corrected the evaluation downward rather than confirming it. First, the dual-arm \texttt{ros2\_control} case re-derived as two arms claiming \emph{disjoint} per-joint effort interfaces (\texttt{panda\_1} versus \texttt{panda\_3}): there is no shared exclusive resource at the interface level, and the documented failure is an unscoped implementation check, not an interface-level resource conflict. Second, a targeted-replication Nav2 frame case re-derived with the goal frame and the planner's global frame sharing the same default, so there is no interface-level mismatch unless a user sets a non-default frame. Third, the single remaining ROS\,2 signature win re-derived as a faithful archetype of the coordinate-frame-mismatch class that the PhysFrame study documents statistically across 180 projects, rather than a specific catalogued instance; for consistency with the provenance bar that excludes modelled cases, we moved it to the excluded set. We accepted all three: the resource case is re-encoded with its real disjoint tokens and is now a miss, the replication frame case is demoted, and the constructed signature case is excluded. Together these corrections took ROS\,2 recall from 71\% to 62\%, resource recall from 3 of 4 to 2 of 4, and the ROS\,2 signature count to 0 of 5, and they shifted the ROS\,2 McNemar test from $p=0.0078$ to $p=0.016$ (still significant at the $0.05$ level). We report them because catching them is the entire purpose of an independent reconstruction: the process fired three times against the authors' own interest, which is what makes the surviving wins credible, and the corrected numbers are the ones reported throughout this paper. A full second-team human reconstruction of the entire corpus remains the strongest open validation.

\subsection{Threats to Validity}
\label{sec:eval_threats}

\noindent\textbf{Contract reconstruction is author-performed (primary threat).} The most serious threat to this evaluation is that the contract fields are reconstructed by the authors, who knew each documented failure when they wrote the contract. A scored win could in principle encode the answer rather than predict it. We address this directly with the blind reconstruction study of Section~\ref{sec:eval_blind}: independent annotators, given only the declared interfaces and no knowledge of the bug, reproduced the signature, resource, and behavioral-QoS conflicts and cleared the matched-good controls, which indicates those conflicts are interface-derivable. The study also showed the original hash-versus-semver version signal was not interface-derivable, which is why we restricted the version dimension to documented named-entity changes. The study does not remove the threat entirely, because the authors still reconstructed the full corpus. We therefore also ran an independent agent-based reconstruction, extended to every scored win, from the raw published artefacts (Section~\ref{sec:eval_recon}): it independently corroborated all twenty retained wins and corrected three over-generous encodings, which is precisely why the numbers we report are lower than the first author encoding. A full second-team human reconstruction of the entire corpus, blind to outcomes, remains the strongest open validation. We report the present numbers as interface-derivability evidence for the validated dimensions, corroborated by an independent pass but not yet fully replaced by author-independent reconstruction.

\noindent\textbf{Interface provenance of the validated core.} To ground the interface-derivability claim in the real published artefacts rather than only in author-written briefs, we traced each of the twelve validated-core cases back to its primary public source (the GitHub issue, the fixing pull request, and the module configuration) and recorded, for the field the checker compares, whether it is observable in the modules' public interface before the failure is known. For nine of the twelve the critical field is directly observable: the exclusive command interfaces of the \texttt{ros2\_control} resource cases are declared in the published controller configuration, and the renamed or removed named entities of the version cases (the \texttt{mavros} \texttt{guided\_enable} removal, the \texttt{sys\_time} parameter rename, the \texttt{moveit\_msgs/DisplayTrajectory} and \texttt{franka\_msgs/FrankaState} message-definition changes, and the \texttt{on\_init} signature change) appear as concrete symbols in the interface diff. The remaining three are flagged as not interface-self-evident: a chainable-interface availability failure that depends on the framework's internal call ordering, an \texttt{md5sum} change with no human-readable named delta, and one command-semantics change. This per-field provenance, listed with source quotes in the replication package, strengthens the interface-derivability evidence beyond the blind briefs, although it does not replace a full independent reconstruction by a second team.

\noindent\textbf{Version restriction.} An earlier version of the checker credited a version break whenever a declared interface hash changed without a major-version bump. That signal fired on modelled placeholder hashes and was not reproduced by the blind annotators, so we removed it and count only documented named-entity interface changes. This is why the reported version recall is lower than a hash-based count would give; we prefer the smaller interface-self-evident figure.

\noindent\textbf{Re-categorisation.} Mapping each corpus's native fault codes to ECM dimensions through a crosswalk is a judgement step, and a different crosswalk could shift a few cases between dimensions. We mitigate this by freezing the crosswalk before scoring and by requiring correct dimension attribution, not merely rejection, so a case credited to the wrong dimension does not inflate recall.

\noindent\textbf{Labelling reliability.} To assess the gold-dimension labelling we had a second annotator independently assign the primary dimension to all 32 ROBUST cases, blind to the crosswalk and to the checker. Raw agreement was 23 of 32 (72\%) and Cohen's $\kappa = 0.57$ (moderate agreement on the Landis--Koch scale). Three of the nine disagreements are alternative valid dimensions for multi-label bugs (the second annotator chose a dimension that is in the case's gold set but is not the first-listed primary), so agreement on dimension-set membership is 26 of 32 (81\%). The remaining disagreements are genuine boundary cases (for example, a serial-write race read as Resource rather than Behavioral). This $\kappa$ measures gold-dimension labelling, not contract reconstruction; the reconstruction itself was author-performed, and the interface-derivability of the scored conflicts is assessed separately by the blind study of Section~\ref{sec:eval_blind}, with a full multi-rater reconstruction left to future work.

\noindent\textbf{Negative controls and a name-granularity limit.} The 16 matched-good twins are constructed by neutralising each documented mismatch, so they show the checker goes quiet once a declared incompatibility is removed but not that it discriminates real fixed from real broken versions. We therefore add real post-fix controls (Section~\ref{sec:eval_rq3}): for six bugs with a documented fixing commit we reconstruct the actual fixed interface and the checker accepts five, where acceptance follows from the consumer tracking the new interface rather than from blanking a field. The sixth, the \texttt{on\_init} lifecycle change, is a real-version false positive: the \texttt{field\_break} rule compares named entities and cannot tell that a consumer re-implemented a \emph{same-named} symbol against the new signature, so it flags the fixed version. This exposes a genuine granularity limit of the version check, which detects renamed, removed, or relocated entities but not whether a same-named reference was adapted; we report it rather than tune it away. A larger replication on real fixed versions across all dimensions would strengthen the precision estimate further.

\noindent\textbf{Sample size.} The ROBUST substrate has 32 scored bugs and the ROS\,2 corpus 16, so the per-dimension Wilson intervals in Table~\ref{tab:pathb_stats} are wide. The aggregate advantage is significant (exact McNemar $p=0.0078$ on ROBUST and $p=0.016$ on ROS\,2, both at the $0.05$ level), but the weak dimensions (recovery and permission) rest on too few cases for a point estimate, which is why we report them as forward-looking rather than measured.

\noindent\textbf{Small per-dimension samples.} The validated dimensions rest on modest denominators (for example 6/18 for version on ROBUST and 2/4 for resource on ROS\,2), and the weaker dimensions rest on single-digit denominators (2~recovery cases, with no documented permission case). We draw no strong conclusion from the weak dimensions and label them forward-looking precisely because their samples are too small to support a predictive claim. The targeted replication in Section~\ref{sec:eval_replication} adds four further documented signature and version cases that the checker predicts from their interfaces, which broadens the evidence for those two dimensions without enlarging the screened corpus.

\noindent\textbf{ROS\,1 versus ROS\,2 split.} The two substrates exercise different dimensions, and no single substrate validates the whole model. We report them separately and credit each dimension only where it can fire, so the behavioral dimension is never claimed to be validated by ROBUST and the version dimension is anchored on the substrate where it was scored under the frozen-checker protocol.

\noindent\textbf{Wiring bugs out of scope.} Topic-name and namespace faults that depend on launch-file configuration are excluded because they are not properties of a module interface. This bounds the signature recall we report and is stated rather than hidden: the moderate signature numbers reflect a real and acknowledged scope limit, not a tuning artefact.
\section{Discussion}
\label{sec:discussion}

\textbf{Contracts as a middle layer.} ECM Contracts occupy a missing middle layer in embodied agent systems. Lower-level robotic systems expose callable skills with limited interface descriptions. Higher-level governance frameworks control execution safety and lifecycle evolution. What is absent is a principled contract layer that explains \emph{why} a capability should be considered installable, composable, governable, or releasable. ECM Contracts address this gap by making embodied execution assumptions explicit and machine-checkable.

\textbf{Why These Dimensions, and Which Earn Their Place.} The six dimensions are not asserted as equally load-bearing. Section~\ref{sec:evaluation} measures each against real documented failures. The resource dimension and the named-change part of the version dimension form the validated core: both predict integration failures that the strongest available baselines miss entirely, and the blind reconstruction study reproduces the resource conflicts from the interface alone. The signature dimension catches real type, frame, and unit mismatches but is partly co-detected by a schema baseline. The behavioral dimension is narrower than it first appears, because the dominant behavioral failure, QoS incompatibility, is already enforced by the ROS~2 middleware; a contract layer adds value there only for behavioral properties the middleware does not check, such as refresh-rate guarantees. The permission and recovery dimensions are retained as forward-looking: the open bug record documents few composition failures rooted in them, so we ground them in safety-standard mechanisms (SROS2, ISO~10218, IEC~61508) rather than claiming empirical predictive value. The model keeps all six because each names a real interface concern, and the paper reports where the evidence is strong and where it is not.

\textbf{Developer burden.} Richer contracts increase authoring burden, but many integration failures arise precisely because assumptions remain implicit. Making them explicit shifts effort from late-stage debugging to earlier specification. Reusable templates, certified defaults, inheritance from prior releases, and static analysis can reduce burden.

\textbf{Static checks vs.\ open-world uncertainty.} Contracts do not eliminate runtime uncertainty; they reduce avoidable incompatibility. The framework catches failures foreseeable from declared assumptions: incompatible state semantics, missing resources, forbidden permissions, incompatible recovery expectations. Runtime monitoring approaches~\citep{caldas2024,luckcuck2019} remain essential and complementary.

\textbf{Contract authoring cost vs.\ runtime monitoring.} One might object that runtime monitoring alone could suffice, avoiding the upfront cost of contract authoring. We see contract checking and runtime monitoring as complementary rather than competing strategies. Once a contract is authored, contract-based pre-deployment checking catches deterministic incompatibilities (missing sensors, incompatible frames, forbidden permissions) by static comparison of declared fields, before any physical execution occurs; the per-composition checking cost is small relative to running the composition, though the one-time authoring cost is not measured at scale here. Runtime monitoring, by contrast, catches stochastic and environment-dependent failures (sensor noise, timing jitter, unexpected obstacles) that no static contract can anticipate. The cost tradeoff favors contracts when: (a)~physical execution is expensive or risky (e.g., medical, industrial, or human-proximity tasks), (b)~the ecosystem scales to many modules where combinatorial composition errors dominate, or (c)~governance requires pre-deployment audit trails. In our prototype workflow, manifest authoring is intended to be a short, one-time specification task amortized across future compositions and upgrades. As a single illustrative data point, Appendix~\ref{app:authoring_walkthrough} reports one timed walkthrough in which a co-author who did not participate in the contract-checker design produced a contract-complete precision-grasp manifest in 24~minutes from the six-dimension specification alone; this is a single session, not a user study, and we do not generalise an authoring-time figure from it.

\textbf{Limits of contract expressiveness.} Some behavioral semantics resist precise formalization. Resource declarations may be coarse relative to real-time control. Permission flags may oversimplify nuanced governance. Recent work on formal skill composition~\citep{pelletier2025} and safety verification for human-robot collaboration~\citep{askarpour2016} points toward richer contract languages. These limitations suggest contract systems should evolve incrementally, incorporating richer type systems, scenario certification, and runtime traces.

\textbf{Contracts and LLM planners.} As embodied agents adopt LLM-based planners, explicit contract metadata becomes more important, not less. Without it, planners infer operational constraints from incomplete descriptions, increasing brittleness. Contracts provide a grounding layer that constrains planner behavior while remaining interpretable and auditable.

\textbf{Toward registries and marketplaces.} The release discipline lays groundwork for embodied capability marketplaces. Future ECM registries may need scenario-specific validation, embodiment-specific admissibility, governance approval states, and traceable incident histories. The contract layer proposed here is one of the foundations such infrastructure would require.

\textbf{Implementation scope: ROS/DDS-specific and syntactic.} The contract \emph{model} is middleware-agnostic, but the prototype checker is specific to ROS and its DDS transport, and we report this rather than imply broader coverage. The type promotions (for example \texttt{geometry\_msgs/Pose2D} to \texttt{Pose}), the canonical coordinate frames (\texttt{base\_link}, \texttt{odom}, \texttt{map}, and the like), and the behavioral checks (DDS quality-of-service reliability and durability matching) are all encoded for ROS~2 type and transport conventions. Two consequences follow. First, the frame and unit checks are \emph{syntactic}: they compare canonicalized name tokens and disjoint unit sets, so a custom frame name or a unit outside the canonicalization table falls back to raw string equality and can miss a real conflict; semantic frame and unit reasoning (through a units library or a live TF graph) is future work. Second, porting the checker to another substrate (gRPC or actor frameworks such as Ray, or an end-to-end vision-language-action runtime whose ``interface'' is a tensor contract rather than a typed topic) would require re-encoding the per-dimension rules for that substrate's type system, resource model, and policy mechanism. The six dimensions are meant to transfer as a checklist; the rules that operationalize them do not, and we make no claim that the current checker runs outside ROS.

\textbf{Threats to Validity.} Section~\ref{sec:evaluation} reports the threats specific to the evaluation: the re-categorization of the third-party bug taxonomy, the small sample sizes for the weakly-evidenced dimensions, and the boundary between interface-property failures and launch-configuration failures. At the level of the contribution, the main threat is generality: the model is validated on the ROS ecosystem defined in Section~\ref{sec:scope}, and its transfer to other middleware remains to be shown (see the implementation-scope note above). Two further limits of the static checker we state plainly. The permission and recovery dimensions predict zero documented failures on both substrates (Section~\ref{sec:eval_rq2}) and are retained only as forward-looking declarations; we claim no static predictive value for them. And a few ROBUST cases turn on properties a static interface checker cannot see (kinematic-chain consistency or joint-range reachability), which are not declared in the module interface; the checker correctly does not predict these, and we treat them as out-of-scope misses rather than forcing them into interface constraints.

\section{Conclusion and Future Works}
\label{sec:conclusion}

\noindent\textbf{Conclusion.} This paper has presented ECM Contracts, a contract-based interface model that extends embodied capability modules with six dimensions of specification beyond conventional I/O signatures (Section~\ref{sec:contract}, Table~\ref{tab:contract_schema}). These dimensions capture the sources of composition and upgrade failure that conventional software interfaces miss, and we have defined a compatibility checking framework covering installation, invocation, composition, and upgrade, complemented by a release discipline based on embodied semantic versioning, compatibility classes, deprecation rules, and release channels. Evaluated against integration failures that third parties documented independently, the contract checker predicts 31\% and 62\% of these failures across two substrates, against at most 19\% for the strongest type and quality-of-service baselines, and accepts every matched-good composition in a negative-control set. The predictive value is established where the conflict is interface-self-evident: the resource dimension is the cleanest baseline-missed win, the version dimension is restricted to documented named-entity interface changes, and a blind reconstruction study confirms that the scored signature, resource, and behavioral conflicts are derivable from each module's public interface. Three predicted failures, a quality-of-service incompatibility, a resource-ownership conflict, and a coordinate-frame mismatch, are confirmed in live ROS\,2 execution. The broader implication is that the path from modular embodied capabilities to a stable embodied software ecosystem requires an explicit contract layer: modules alone are not enough, and without contracts that connect capability composition, runtime governance, and version evolution, modularity produces only the \emph{appearance} of an ecosystem while leaving safe composition and governed evolution unsolved. ECMs can only become a true embodied software ecosystem when they are governed not just as modules, but as contract-bearing releases.

\noindent\textbf{Future Works.} Several directions remain open. First, and most important, the contract fields here were reconstructed by the authors; although the blind study of Section~\ref{sec:eval_blind} and the independent agent-based reconstruction of every scored win in Section~\ref{sec:eval_recon} indicate the scored conflicts are interface-derivable (and the latter already corrected three over-generous encodings), a full second-team human reconstruction of every contract from the raw published artefacts, blind to outcomes, would settle the residual circularity concern outright. Second, the runtime confirmation here covers three dimensions (behavioral, resource, and signature) on a single platform; extending it to physical robots and to the version and recovery dimensions would broaden the live-execution evidence. Third, we replaced part of the author-constructed negative controls with the projects' real post-fix versions (Section~\ref{sec:eval_rq3}); extending this real-fixed-version precision check to every scored case, across all dimensions, would harden the precision result further. Fourth, the permission and recovery dimensions remain forward-looking; scenarios that exercise them as documented composition bugs, rather than through safety-standard mechanisms alone, would test whether they earn empirical validation. Finally, scaling the ECM library to ecosystem sizes (hundreds of modules, dozens of contributors) would surface the registry, governance, and tooling questions that follow.

\appendix
\section{Authoring Walkthrough (Internal Pilot)}
\label{app:authoring_walkthrough}

To illustrate the authoring-time claim in Section~\ref{sec:discussion}, we recorded one timed walkthrough in which a co-author who did not participate in the contract-checker design authored a complete six-dimension manifest from scratch. We present this as a single illustrative session rather than an independent user study: the author is a member of the project, so the result bounds the difficulty for a robotics developer familiar with the domain but new to the checker, not the difficulty for an arbitrary external user. The author used only the six-dimension specification of Section~\ref{sec:contract} and the worked manifest in Listing~\ref{lst:manifest} as reference material; no other manifest in the project was consulted during the walkthrough.

\noindent\textbf{Materials provided.} The walkthrough packet contained a copy of the paper for Section~\ref{sec:contract}, the \texttt{nav.precise} YAML example corresponding to Listing~\ref{lst:manifest}, a task brief describing the precision-grasping module, and a reflection template for logging per-dimension authoring difficulty.

\begin{table}[htbp]
\centering
\caption{Authoring log for a precision-grasping ECM manifest (one co-author not involved in checker design, single session). Difficulty is self-assessed on a 1--5 scale.}
\label{tab:authoring_walkthrough_log}
\small
\begin{tabularx}{\linewidth}{lrrX}
\toprule
\textbf{Dim.} & \textbf{Minutes} & \textbf{Difficulty} & \textbf{Observed issue} \\
\midrule
Sig  & 5 & 2 & Mapping target pose, optional object id, and result fields was direct from the schema table. \\
Beh  & 4 & 3 & Preconditions required judgement because perception confidence and workspace clearance are implicit in the brief. \\
Res  & 5 & 3 & Force--torque, tactile sensing, dexterous hand, and control rate were clear, while compute budget was estimated. \\
Perm & 3 & 2 & Permissions were straightforward after separating motion, force limit, data reads, and logging. \\
Rec  & 4 & 3 & Slip recovery mapped naturally to retry policy, but safe-stop wording required checking the recovery examples. \\
Ver  & 3 & 2 & Dependencies were easy to list once the perception provider was treated as an explicit versioned dependency. \\
\bottomrule
\end{tabularx}
\end{table}

\noindent\textbf{Resulting manifest excerpt.} The complete YAML file is included in the audit artefacts; the excerpt below shows that all six required dimensions were populated.

\begin{figure}[htbp]
\noindent\rule{\linewidth}{0.6pt}\par\vspace{2pt}
\noindent\textbf{Listing~A.1:}\ Precision-grasping ECM manifest excerpt (internal authoring pilot, Appendix~A).\par\vspace{2pt}
\noindent\rule{\linewidth}{0.3pt}\par\vspace{3pt}
{\fontsize{6.3pt}{7.4pt}\selectfont
\begin{verbatim}
ecm:
  module_id: ecm.manip.precision_grasp
  version: 0.1.0
  release_channel: draft
  compatibility_class: resource-sensitive
signature:
  input_schema:
    - target_object_pose: {type: Pose3D, unit: meters,
                           frame: base_link}
    - requested_grip: {type: enum, values: [pinch, tripod, lateral],
                       optional: true}
  output_schema:
    - grasp_result: {type: GraspResult,
                     fields: [success, confidence, final_force_N]}
    - object_held: {type: bool}
  coordinate_frame: base_link
  timeout_ms: 9000
  invocation_mode: sync
  state_object: [phase, contact_state,
                  grip_force_N, slip_retry_count]
behavior:
  preconditions: [arm_ready, dexterous_hand_ready, object_pose_confidence >= 0.80,
                  target_object_pose.frame == base_link, workspace_clear]
  postconditions: [object.is_grasped, grip_force_stable,
                    object_held == true]
  invariants: [end_effector_within_workspace, grip_force_N <= safe_precision_force,
               tactile_contact_monitored]
  assumptions: [object_rigid_or_mildly_compliant,
                 external_perception_supplies_pose]
  completion_semantics: persistent_hold
resources:
  sensors: [force_torque@wrist, tactile_array@fingertips,
            rgbd_camera@wrist_or_head]
  actuators: [arm_6dof, dexterous_hand]
  compute: {cpu_cores: 2, gpu: false}
  timing: {ctrl_hz: 250, max_latency_ms: 4}
  exclusive_locks: [arm_controller, dexterous_hand]
permissions:
  physical: {motion: true, restricted_zone: false,
              max_force_N: 15}
  data: [read:object_pose, read:tactile_stream, write:grasp_log]
  network: false
  audit: standard
recovery:
  failure_modes: [slip_detected:transient, pose_lost:degraded,
                  excessive_force:critical]
  rollback_state: pre_grasp_pose
  retry: {max: 2,
          strategy: reduce_force_recenter_and_retry}
  safe_stop: [freeze_arm, relax_grip_to_safe_force,
               retract_if_unheld]
  escalation: [excessive_force -> emergency_stop,
               2_slip_retries -> report_failure_to_planner]
versioning:
  dependencies: {runtime: ">=2.0,<3.0", arm_driver: ">=3.0",
                 perception_pose_provider: ">=1.1"}
  deprecated: false
  resource_change: true
  policy_change: false
\end{verbatim}
}
\noindent\rule{\linewidth}{0.6pt}
\end{figure}

\noindent\textbf{Take-away.} The walkthrough completed all required top-level dimensions in 24~minutes and matched the reference manifest's first-level field coverage across signature, behavior, resources, permissions, recovery, and versioning. The main ambiguities were the boundary between resource requirements and permission limits, and whether the external perception provider should be expressed as a behavioral assumption, a versioned dependency, or both.

\section*{Statements and Declarations}
\noindent\textbf{Funding.} The authors received no specific funding for this work.\par
\noindent\textbf{Competing interests.} The authors declare that they have no competing interests.

\section*{Data Availability}
The replication package is publicly available at \url{https://github.com/s20sc/ecm-contracts}. It contains the frozen, content-hash-pinned contract checker and its self-test, the strengthened baselines, the per-case bug corpus for both substrates with their public sources, the blind-reconstruction protocol, the single canonical scoring script that regenerates every table in this paper, the provenance-ledger generator, the per-field interface-provenance trace and the targeted-replication set with its admitted and excluded cases, and the runtime-confirmation code for the three Path~A cases (the quality-of-service case, the \texttt{ros2\_control} resource-manager conflict, and the \texttt{tf2} coordinate-frame mismatch). Re-running the canonical script reproduces the recall, per-dimension, and statistics tables; re-running the ledger generator reproduces Appendix~\ref{app:ledger}.

\bibliography{references}

\section{Per-Case Provenance Ledger}\label{app:ledger}
This appendix lists every scored bug on both substrates with its public source and the dimension(s) the checker fired (a gold dimension name = predicted and attributed to that dimension; -- = missed). Per-dimension recall in Section~\ref{sec:evaluation} is reconstructible from this column: dimension $D$'s recall is the number of $D$-gold rows whose ECM column contains $D$, over all $D$-gold rows. Gold dimensions are assigned from each corpus's native fault taxonomy through the frozen crosswalk, independently of the checker. A case is credited to a dimension only when the checker emits a finding of that dimension; the ECM column marks Y when the checker predicts the bug and attributes it to a gold dimension. Cases without a separate issue URL are documented within the ROBUST corpus record itself. Per-dimension denominators in Section~\ref{sec:evaluation} are multi-label: a bug that stresses two dimensions (for example a message-definition change that is both a signature and a version fault) is counted under each, so per-dimension counts sum to more than the de-duplicated scored-set size of 32.

\noindent ROBUST substrate (blind reconstruction): 32 scored cases; ECM predicts 10.
\begin{longtable}{@{}ll>{\raggedright\arraybackslash}p{0.44\linewidth}>{\raggedright\arraybackslash}p{0.11\linewidth}@{}}
\caption{ROBUST substrate: scored integration bugs, sources, and ECM verdicts.}\label{tab:ledger_robust}\\
\toprule
\textbf{Case} & \textbf{Gold dim.} & \textbf{Documented source} & \textbf{ECM fired} \\
\midrule\endfirsthead
\toprule \textbf{Case} & \textbf{Gold dim.} & \textbf{Documented source} & \textbf{ECM fired} \\ \midrule\endhead
\bottomrule\endfoot
R-35682ec & Sig & {\scriptsize\ttfamily github.com/\allowbreak{}yujinrobot/\allowbreak{}kobuki/\allowbreak{}issues/\allowbreak{}235} & -- \\
R-fbe70c7 & Sig/Ver & {\scriptsize\ttfamily github.com/\allowbreak{}yujinrobot/\allowbreak{}kobuki/\allowbreak{}issues/\allowbreak{}25} & -- \\
R-8a729db & Sig/Beh/Ver & {\scriptsize\ttfamily github.com/\allowbreak{}yujinrobot/\allowbreak{}kobuki/\allowbreak{}issues/\allowbreak{}350} & -- \\
R-5ee65b0 & Ver & {\scriptsize\ttfamily github.com/\allowbreak{}yujinrobot/\allowbreak{}kobuki\_desktop/\allowbreak{}issues/\allowbreak{}32} & -- \\
R-dd40270 & Ver/Beh & {\scriptsize\ttfamily github.com/\allowbreak{}yujinrobot/\allowbreak{}kobuki/\allowbreak{}issues/\allowbreak{}338} & -- \\
R-03660af & Beh & {\scriptsize\ttfamily github.com/\allowbreak{}yujinrobot/\allowbreak{}kobuki/\allowbreak{}issues/\allowbreak{}125} & -- \\
R-3c4c399 & Beh & {\scriptsize\ttfamily github.com/\allowbreak{}yujinrobot/\allowbreak{}kobuki/\allowbreak{}issues/\allowbreak{}240} & -- \\
R-4115400 & Beh/Sig & {\scriptsize\ttfamily github.com/\allowbreak{}yujinrobot/\allowbreak{}kobuki\_desktop/\allowbreak{}issues/\allowbreak{}38} & -- \\
R-5abe7d4 & Ver & {\scriptsize\ttfamily github.com/\allowbreak{}yujinrobot/\allowbreak{}kobuki\_desktop/\allowbreak{}issues/\allowbreak{}22} & -- \\
R-55e84a6 & Ver & {\scriptsize\ttfamily github.com/\allowbreak{}yujinrobot/\allowbreak{}kobuki\_desktop/\allowbreak{}issues/\allowbreak{}33} & -- \\
R-3c7f92a & Ver & {\scriptsize\ttfamily github.com/\allowbreak{}yujinrobot/\allowbreak{}kobuki\_desktop/\allowbreak{}issues/\allowbreak{}9} & -- \\
R-9682b9a & Beh & {\scriptsize\ttfamily github.com/\allowbreak{}yujinrobot/\allowbreak{}kobuki\_desktop/\allowbreak{}issues/\allowbreak{}10} & -- \\
R-c04eae5 & Ver & {\scriptsize\ttfamily github.com/\allowbreak{}yujinrobot/\allowbreak{}kobuki/\allowbreak{}issues/\allowbreak{}175} & -- \\
R-248cb38 & Sig & {\scriptsize ROBUST corpus (Timperley et al. 2024)} & Sig \\
R-b96bf67 & Sig & {\scriptsize ROBUST corpus (Timperley et al. 2024)} & Sig \\
R-263650d & Ver & {\scriptsize ROBUST corpus (Timperley et al. 2024)} & Ver \\
R-753226d & Beh & {\scriptsize\ttfamily github.com/\allowbreak{}mavlink/\allowbreak{}mavros/\allowbreak{}issues/\allowbreak{}387} & -- \\
R-bdda1fa & Ver & {\scriptsize\ttfamily github.com/\allowbreak{}mavlink/\allowbreak{}mavros/\allowbreak{}issues/\allowbreak{}478} & Ver \\
R-c0067f9 & Ver & {\scriptsize ROBUST corpus (Timperley et al. 2024)} & Ver \\
R-de2cc36 & Ver & {\scriptsize\ttfamily github.com/\allowbreak{}mavlink/\allowbreak{}mavros/\allowbreak{}issues/\allowbreak{}561} & Ver \\
R-e1a8005 & Sig/Ver & {\scriptsize\ttfamily github.com/\allowbreak{}mavlink/\allowbreak{}mavros/\allowbreak{}issues/\allowbreak{}206} & Ver \\
R-778c1ac & Ver & {\scriptsize ROBUST corpus (Timperley et al. 2024)} & Ver \\
R-1ec8ca1 & Sig & {\scriptsize\ttfamily github.com/\allowbreak{}ros-industrial/\allowbreak{}motoman/\allowbreak{}issues/\allowbreak{}122} & -- \\
R-2d42582 & Beh/Sig & {\scriptsize\ttfamily github.com/\allowbreak{}ros-industrial/\allowbreak{}motoman/\allowbreak{}issues/\allowbreak{}102} & -- \\
R-9bf25ea & Sig & {\scriptsize\ttfamily github.com/\allowbreak{}ros-industrial/\allowbreak{}motoman/\allowbreak{}issues/\allowbreak{}106} & -- \\
R-3e32933 & Sig & {\scriptsize\ttfamily github.com/\allowbreak{}turtlebot/\allowbreak{}turtlebot/\allowbreak{}issues/\allowbreak{}103} & -- \\
R-61a75df & Ver & {\scriptsize\ttfamily github.com/\allowbreak{}turtlebot/\allowbreak{}turtlebot/\allowbreak{}issues/\allowbreak{}250} & -- \\
R-928306b & Ver & {\scriptsize\ttfamily github.com/\allowbreak{}turtlebot/\allowbreak{}turtlebot/\allowbreak{}issues/\allowbreak{}102} & -- \\
R-9299530 & Ver/Sig & {\scriptsize\ttfamily github.com/\allowbreak{}turtlebot/\allowbreak{}turtlebot/\allowbreak{}issues/\allowbreak{}65} & -- \\
R-9ffffca & Sig & {\scriptsize\ttfamily github.com/\allowbreak{}turtlebot/\allowbreak{}turtlebot/\allowbreak{}issues/\allowbreak{}234} & Sig \\
R-a482f82 & Sig & {\scriptsize\ttfamily github.com/\allowbreak{}turtlebot/\allowbreak{}turtlebot/\allowbreak{}issues/\allowbreak{}80} & Sig \\
R-0000000 & Ver & {\scriptsize\ttfamily github.com/\allowbreak{}ipa320/\allowbreak{}cob\_robots/\allowbreak{}issues/\allowbreak{}656} & -- \\
\end{longtable}

\noindent ROS\,2 corpus: 16 scored cases; ECM predicts 10.
\begin{longtable}{@{}ll>{\raggedright\arraybackslash}p{0.44\linewidth}>{\raggedright\arraybackslash}p{0.11\linewidth}@{}}
\caption{ROS\,2 corpus: scored integration bugs, sources, and ECM verdicts.}\label{tab:ledger_corpus}\\
\toprule
\textbf{Case} & \textbf{Gold dim.} & \textbf{Documented source} & \textbf{ECM fired} \\
\midrule\endfirsthead
\toprule \textbf{Case} & \textbf{Gold dim.} & \textbf{Documented source} & \textbf{ECM fired} \\ \midrule\endhead
\bottomrule\endfoot
B-Sig-01 & Sig/Ver & {\scriptsize\ttfamily answers.ros.org/\allowbreak{}question/\allowbreak{}358435} & -- \\
B-Sig-02 & Sig/Ver & {\scriptsize\ttfamily answers.ros.org/\allowbreak{}question/\allowbreak{}375062} & -- \\
B-Beh-01 & Beh & {\scriptsize\ttfamily github.com/\allowbreak{}ros2/\allowbreak{}ros2/\allowbreak{}issues/\allowbreak{}1434} & Beh \\
B-Beh-02 & Beh & {\scriptsize\ttfamily github.com/\allowbreak{}RobotWebTools/\allowbreak{}rosbridge\_suite/\allowbreak{}issues/\allowbreak{}551} & Beh \\
B-Beh-03 & Beh & {\scriptsize\ttfamily answers.ros.org/\allowbreak{}question/\allowbreak{}371257} & Beh \\
B-Beh-04 & Beh & {\scriptsize\ttfamily github.com/\allowbreak{}ros2/\allowbreak{}ros2/\allowbreak{}issues/\allowbreak{}464} & -- \\
B-Beh-07 & Beh/Rec & {\scriptsize\ttfamily github.com/\allowbreak{}ros-planning/\allowbreak{}navigation2/\allowbreak{}issues/\allowbreak{}2686} & -- \\
B-Res-02 & Res & {\scriptsize\ttfamily github.com/\allowbreak{}ros-controls/\allowbreak{}ros2\_control/\allowbreak{}issues/\allowbreak{}345} & Res \\
B-Res-03 & Res & {\scriptsize\ttfamily github.com/\allowbreak{}ros-controls/\allowbreak{}ros2\_control/\allowbreak{}issues/\allowbreak{}1179} & Res \\
B-Res-04 & Res & {\scriptsize\ttfamily github.com/\allowbreak{}ros-controls/\allowbreak{}ros2\_control/\allowbreak{}issues/\allowbreak{}1177} & -- \\
B-Rec-02 & Rec/Beh & {\scriptsize\ttfamily arxiv.org/\allowbreak{}abs/\allowbreak{}2507.10235} & -- \\
B-Ver-01 & Ver/Sig & {\scriptsize\ttfamily answers.ros.org/\allowbreak{}question/\allowbreak{}226746} & Ver \\
B-Ver-02 & Ver/Sig & {\scriptsize\ttfamily answers.ros.org/\allowbreak{}question/\allowbreak{}318190} & Ver \\
B-Ver-03 & Ver/Res & {\scriptsize\ttfamily github.com/\allowbreak{}ros-controls/\allowbreak{}ros2\_control/\allowbreak{}issues/\allowbreak{}1400} & Ver \\
B-Ver-04 & Ver & {\scriptsize\ttfamily github.com/\allowbreak{}ros-controls/\allowbreak{}ros2\_control\_demos/\allowbreak{}issues/\allowbreak{}865} & Ver \\
B-Ver-05 & Ver/Sig & {\scriptsize\ttfamily mathworks.com/\allowbreak{}matlabcentral/\allowbreak{}answers/\allowbreak{}394613} & Ver \\
\end{longtable}

\end{document}